\documentclass[journal]{IEEEtran}
\usepackage[T1]{fontenc}
\usepackage[latin9]{inputenc}
\usepackage{mathtools}
\usepackage{amsmath}
\usepackage{amssymb}
\usepackage{graphicx}
\usepackage{colortbl}

\makeatletter

\IEEEoverridecommandlockouts
\usepackage{array}
\usepackage{soul}
\usepackage{xcolor}
\usepackage{tikz}
\usepackage{textcomp}

\definecolor{100}{HTML}{FF0000}%
\definecolor{89}{HTML}{FF8000}%
\definecolor{47}{HTML}{FFBF00}%
\definecolor{11}{HTML}{FFFF00}%
\definecolor{5}{HTML}{BFFF00}%
\definecolor{3}{HTML}{08FF00}%

\makeatother

\begin{document}
\pagestyle{empty}

%\IEEEoverridecommandlockouts\IEEEpubid{}

\title{A Multisensory Edge-Cloud Platform for Opportunistic Radio Sensing
in Cobot Environments}
\author{Sanaz Kianoush,~\IEEEmembership{member,~IEEE}, Stefano Savazzi,~\IEEEmembership{member,~IEEE},
Manuel Beschi, Stephan Sigg,~\IEEEmembership{member,~IEEE}, and
Vittorio Rampa,~\IEEEmembership{member,~IEEE} \thanks{S. Kianoush, S. Savazzi and V. Rampa are with the Institute of Electronics,
Computer and Telecommunication Engineering (IEIIT) of Consiglio Nazionale
delle Ricerche (CNR), p.zza Leonardo da Vinci, 20133 Milano, Italy,
e-mail: \{sanaz.kianoush, stefano.savazzi, vittorio.rampa\}@ieiit.cnr.it.} \thanks{M. Beschi is with the Institute of Sistemi e Tecnologie Industriali
Intelligenti per il Manifatturiero Avanzato (STIIMA), of Consiglio
Nazionale delle Ricerche (CNR), Milano, Italy, e-mail: \{manuel.beschi@stiima.cnr.it.}\thanks{Stephan Sigg is with Aalto University, Department of Communications
and Networking, Finland, e-mail: \{stephan.sigg\}@aalto.fi.} \thanks{This work has been partially funded by MUR Italy and Academy of Finland within the ERA-NET CO-FUND H2020 CHISTERA III project RadioSense.} }

\maketitle
\thispagestyle{empty} %\copyrightnotice % As a general rule, do not put math, special symbols or citations

\begin{abstract}
Worker monitoring and protection in collaborative robot (cobots) industrial
environments requires advanced sensing capabilities and flexible solutions
to monitor the movements of the operator in close proximity of moving
robots. Collaborative robotics is an active %represents a mushrooming 
research area where Internet of Things (IoT) and novel sensing technologies are
expected to play a critical role. Considering that no single technology
can currently solve the problem of continuous worker monitoring, the
paper targets the development of an IoT multisensor data fusion (MDF) platform.
It is based on an edge-cloud architecture that supports the combination 
and transformation of multiple sensing technologies to enable the passive 
and anonymous detection of workers.
Multidimensional data acquisition from different IoT sources, signal
pre-processing, feature extraction, data distribution and fusion,
along with machine learning (ML) and computing methods are described. The
proposed IoT platform also comprises a practical solution for data
fusion and analytics. It is able to perform opportunistic and real-time perception
of workers by fusing and analyzing radio signals obtained
from several interconnected IoT components, namely a multi-antenna WiFi installation (2.4-5 GHz), a sub-THz imaging camera (100 GHz), a 
network of radars (122 GHz) and infrared sensors (8-13~$\mathbf{\mu}$m). 
The performance of the proposed IoT platform is validated through real 
use case scenarios inside a pilot industrial plant in which protective 
human--robot distance must be guaranteed considering latency and 
detection uncertainties. 
\end{abstract}



% \fi

\IEEEpeerreviewmaketitle{}

\section{Introduction}

\IEEEPARstart{N}{ext} generation manufacturing, namely Industry 4.0
(I4.0), must ensure workspace safety and production efficiency during
human--robot cooperation (HRC). Within this context, human safety
must be guaranteed by continuously and accurately sensing and tracking worker activities. 
Perception of workers in highly dynamic environments thus becomes the key to achieve a high level of efficiency and flexibility
that is required by the foreseen I4.0 applications~\cite{hrc,hrc-wearable}.
Dynamic collaborative robot (cobot) environments entail reconfigurable
layouts, high degrees of shared resources, and humans moving in loosely
structured, fenceless workplaces~\cite{vic} interacting with multiple
machines~\cite{hrc2}. Collaborative robotics is a fairly unique
and challenging application where Internet of Things (IoT) technologies
are expected to play a pivotal role. In this context, the reuse, or
transformation~\cite{transf}, of sensors~\cite{hrc-vision} and
their orchestration through reliable and low-latency machine-type
communications~\cite{commag_2017} (5G wireless technologies~\cite{5G-chapter})
represent the principal components for more integrated, adaptive,
configurable and, above all, safer manufacturing environment.

\begin{figure}
\includegraphics[width=0.48\textwidth]{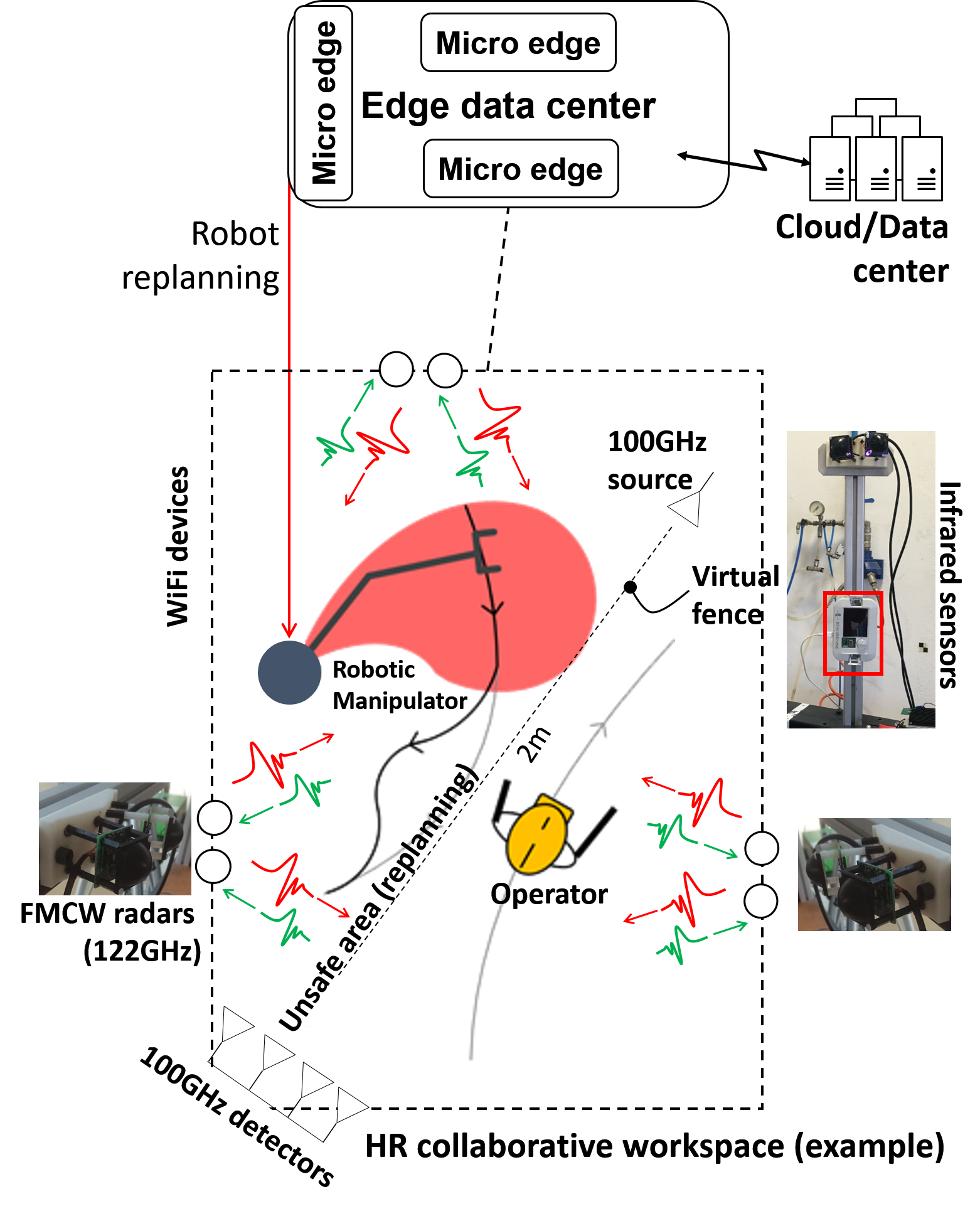} \caption{Multisensor platform and 
collaborative HR workspace consisting of a robotic manipulator and a worker 
in a fenceless shared space.}
\label{figureTeaser-1} 
\end{figure}

Recently, techniques to capture and process the wireless stray electromagnetic
(EM) radiations originated from different radio sources are gaining
increasing attention~\cite{magazine2017,mag2016}. In particular,
these techniques can be exploited for privacy-preserving human-scale
sensing~\cite{iot22}, human behavior recognition, detection/localization,
as well as crowd density estimation and mapping. For example, passive
RF sensing, or radio vision~\cite{mag2016}, leverages different radio
frequency (RF) technologies for sensing tasks. RF signals are perturbed
by moving objects/bodies, and by changing configurations,
due to the propagation of transmitted, reflected, scattered and diffracted EM waves.
Hence, in addition to transporting modulated information, they can
serve as virtual sensors to infer two- or three-dimensional (2D/3D)
views of all objects traversed by the EM wavefield.

Previous work on sensing for HRC usually focused the design of individual technologies, such as wearables~\cite{hrc-wearable},
vision~\cite{hrc-vision}, radars~\cite{sub-TH_2019}, or machine
type communications~\cite{computer2019}. However, it is expected
that the combination and transformation of multiple sensing technologies
will be the key to meet the expected localization accuracy in cobot
environments~\cite{hrc-vision}. The paper proposes the integration
of heterogeneous RF sensing technologies into an industry-compliant
IoT-based edge-cloud computing architecture to provide augmented information
about worker safety in the context of HRC manufacturing. Fig.~\ref{figureTeaser-1}
shows a schematic of the proposed IoT Multisensor Data Fusion (MDF) platform inside
a robotic cell consisting of an industrial manipulator and a human
worker, along with an edge-cloud computing architecture for data distribution
and processing.
The robotic cell implements the closed-loop control of the manipulator
activities to guarantee worker safety. This control loop is underpinned
by the proposed human sensing tool: accuracy and latency tradeoff
is analyzed to achieve safety monitoring of the workspace, namely
to minimize the human-robot protective distance. 

\section{Literature review and contributions}

Opportunistic sensing targets the cross-fertilization of computing
and communication technologies, leveraging opportunistically different
ambient signals (\emph{e.g.}, radio, acoustic, light). The compound effect
of heterogeneous sensors is expected to improve accuracy and to enable
new capabilities. Technologies for opportunistic and transformative
radio sensing, such as  device-free radio localization, activity recognition~\cite{wang2016human},
and people counting~\cite{win}, typically focus on the augmentation and
transformation of existing radio devices, such as WiFi, machine-type
communication (MTC) or cellular-wireless wide area networks (WWAN)~\cite{cellsavazzi2017}
into human-scale sensors. These radio technologies generally exploit
electromagnetic (EM) fields maintained by different sources (\emph{i.e.},
infrared, micro-wave and~$100-122$ GHz, bands). Measurements of such
EM fields are used to extract an image of the environment, or its
modifications, for various motion perception tasks~\cite{sub-TH_2019}.
In particular, the human presence in the environment changes the EM
propagation characteristics due to reflection, scattering and diffraction. 
In~\cite{win}, WiFi signals are used for target counting and activity recognition while, in~\cite{IoT_2018}, a cloud-IoT platform is proposed to sustain  device-free human sensing by real time processing of the channel quality information (CQI). 
In order to manage high-dimensional data processing and real-time analysis, a cloud platform along with machine learning tools are proposed. Multidimensional data analytics is therein implemented by exploiting WiFi signals, while the problems related to data fusion, edge-side computations and service orchestration are not discussed. In~\cite{IoT2016}, a  device-free localization 
and fall detection system has been proposed for HRC applications. It adopts
unmodified machine type radio communications based on the IEEE 802.15.4
standard.

All above mentioned works focus on the evaluation of individual technologies
for human sensing; however, they also demonstrate that no single sensing
technology can provide robust and accurate sensing information.

\subsection{Human-Robot Cooperation and Sensors}

\label{subsec:hrc_cooperation}
In recent years, industrial robotic cells have been subjected to a paradigm shift, where physical fences have been removed to save space and to
increase the interaction between workers and robots~\cite{euroc_book}.
HRC is a key enabler in advanced manufacturing, allowing the human workers to be assisted by cobots for collaborative execution of complex
repetitive workflows. 
Effectiveness in HRC tasks largely depends on the possibility to allow a human operator a great autonomy in decision making, task execution order and timing. 
Examples include, but are not limited to holding/resuming some tasks, temporarily leaving workstations
for contingencies, buffering and re-sequencing production steps~\cite{HALME2018111}.

Advanced HRC tasks involving worker co-presence will be underpinned
by IoT technologies supporting new human-scale monitoring and perception
tools. Focusing on human sensing in the context of cobot environments,
the standard ISO/TS 15066~\cite{ISOTS15066} describes the specifications
of the speed and separation monitoring (SSM) collaborative operations. 
SSM allows robots and workers to share the same workspace while the 
robot has to reduce its speed or/and to follow an alternative path 
according to the worker's position. An effective implementation
of SSM requires the online detection of the relative distance~$d$
between the robot and the worker. This distance is time-sensitive and must
be compared in real-time with the \emph{protective separation distance}
$d_{p}$, namely the minimum permissible human-robot distance. In
turn, the protective distance depends on the uncertainties of the
positions of both robot and human, as well as the relative velocity
between the worker and the robot, and the latencies of the system. The protective distance
$d_{p}$ must be minimized to guarantee an efficient collaborative
space. In addition, the constraint~$d>d_{p}$ must be verified during
all collaborative activities. As analyzed in detail in Sect.~\ref{sec:Protective-human--robot-distance},
the protective distance~$d_{p}$ depends on the reaction time, latency
and positioning uncertainties that should be quantified according
to the deployed sensing system.

In what follows, we refer to localization in terms of \emph{anonymous}
(\emph{i.e.}, passive, or device-free) detection of motion and location, 
as opposed to \emph{active} tracking where identity tagging 
occurs~\cite{hrc-wearable}. Localization
services enabled by the proposed IoT MDF platform should
be able to map increasingly complex figures such as: 
\begin{itemize}
\item \emph{pose}, namely application-space current location coordinates of workers (and cobots); 
\item \emph{trajectory}: first/second order kinematics used for tracing
the projection of the current pose and predicting space occupancy; 
\item \emph{behaviors}: full trajectory patterns associated with some specific
worker activity and semantics. 
\end{itemize}
Pose and trajectory estimation are here explored in detail as necessary for maintaining the protective distance~$d_{p}$. Behavior detection is used to anticipate potentially hazardous situations.
Considering that the effectiveness of pose and trajectory estimation
depends on latency and localization uncertainties, we explore the
compound effect of multiple passive radio sensing technologies with
the goal of optimizing~$d_{p}$.

Mainstream safety technologies make use of dedicated sensors to detect presence, tracing displacements or distances. 
Optoelectronic devices~\cite{hrc-vision} typically exploit the properties of reflectivity, while other technology uses emissivity of thermal images or acoustic fields. The major downsides of these systems are the full-accuracy ranges, the presence of occlusions, the environmental conditions (\emph{e.g.}, dust, fumes) and, for those based on vision tracking, privacy concerns. 
Orchestration of these sensors, through collaborative approaches and data fusion systems based on machine learning, is a critical problem and still
considered open~\cite{hrc3}. 
Besides cooperation of interconnected sensors, transformative, or opportunistic sensing~\cite{transf}
adapts the individual IoT devices to support advanced tasks, far beyond
their original designs.

\begin{figure*}
	\centering \includegraphics[width=0.8\textwidth]{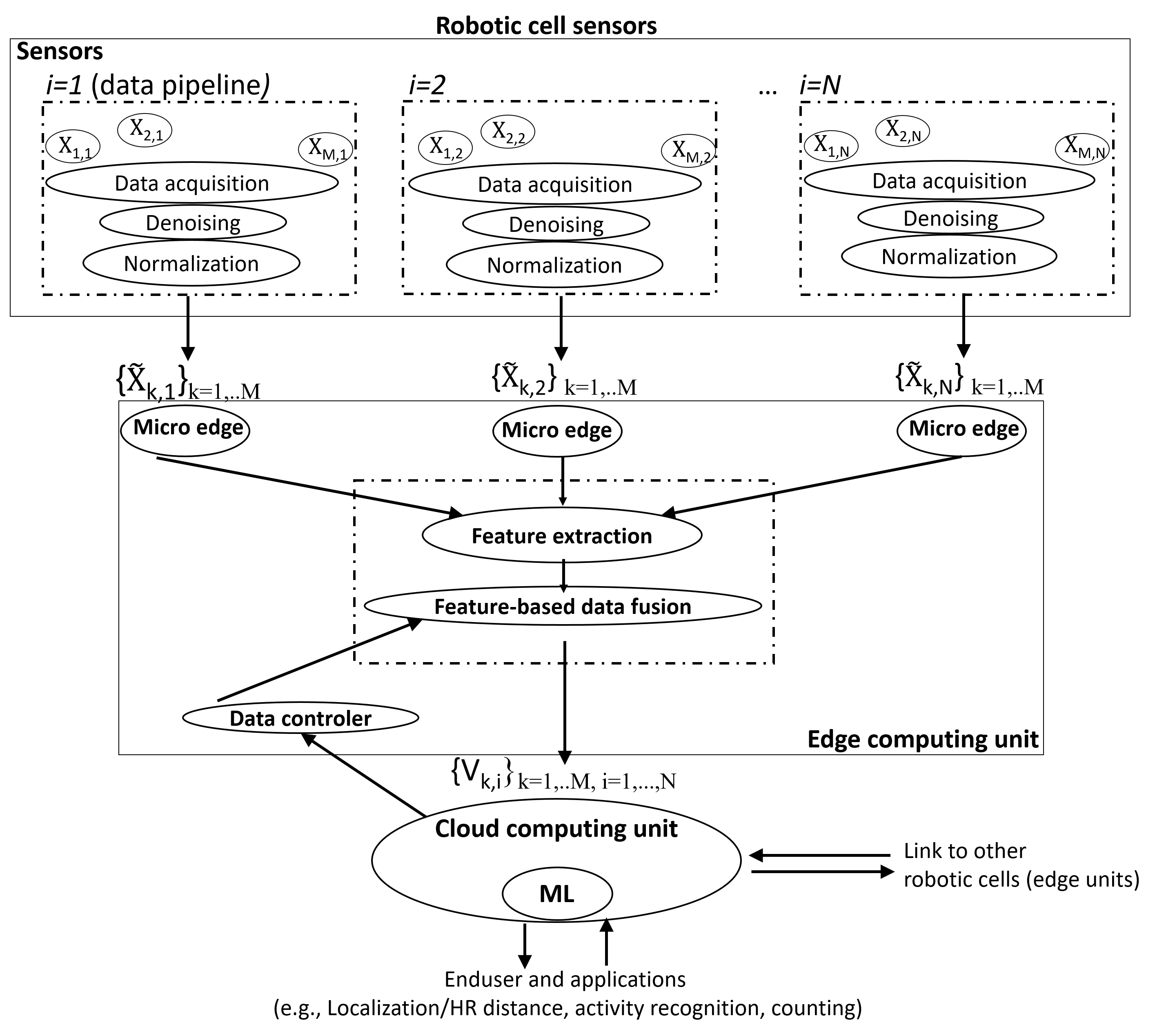}
	\caption{Multisensor data fusion system model and data analytics}
	\label{fig-edge-cloud} 
\end{figure*}

\subsection{Contributions}

\label{subsec:Contributions}

The paper targets the design of an ecosystem of heterogeneous IoT radio-based systems that combines and transforms multiple RF sources/detectors into different
\emph{virtual} sensors. These sensing modalities are all instrumental
to workspace monitoring and enable three key \emph{HRC functions}:
\emph{i)} the identification of the number of workers present in the
workspace, \emph{ii)} the localization of the worker inside the operating
space ($d>1$~m) and \emph{iii)} the monitoring of the HR distance
inside the collaborative space ($d<1$ m). As depicted in Fig.~\ref{figureTeaser-1}, different passive radio sensing IoT devices cooperate to achieve opportunistic and anonymous perception of the worker when sharing the space with a robot. In particular, the proposed MDF platform 
combines IoT sensors that detect infrared (IR) body emissions in 
the short-range, multi-antenna WiFi radios that monitor body-induced 
reflections, a network of radars monitoring the sub-THz ($122$ GHz)
band, and a sub-THz imaging source/camera ($100$ GHz). Cooperation and 
coordination of heterogeneous sensors is based on a cloud-edge platform. 
Considering that multiple radios and sensing sources are employed, we also 
propose a practical feature-based solution for data fusion and analytics. 
Features, namely uniform statistical representations of raw data from 
different sensors, are obtained by edge units. These units are responsible 
for pre-processing and data fusion. Analytics is carried out 
in the cloud and it is based on the processing of the features published 
by the edge units. The proposed cloud-edge MDF platform is designed to 
process large amounts of (multi-dimensional) raw data~\cite{IoT_2018} 
and adopts different machine learning (ML) tools that expect individual, or combinations of, features as input, depending on the HRC function.

The paper is organized as follows: the cloud-edge MDF platform is 
introduced in Sect.~\ref{sec:Multisensory-edge-cloud-platform},
considering feature extraction from pre-processed raw data, feature 
fusion implemented on a dedicated edge device, analytics on the cloud 
back-end, and related communication
aspects, respectively. In particular, the edge node consists of multiple
micro servers/data centers (micro-edges) for extracting, processing
and fusing features from different sensors (pipelines). Focusing on
an HRC workspace scenario where the edge node is configured to monitor
an individual robotic cell (Fig.~\ref{figureTeaser-1}), the IoT components,
sensors and data pre-processing tools are discussed in Sect.~\ref{architect}.
Extensive experimental activities inside the robotic cell have been
conducted in Sect.~\ref{Experiment} to assess the feasibility and
performance of the platform considering the three HRC functions previously
mentioned. The goal is to find the best performance trade-off in terms
of fused data/features volume (\emph{i.e.}, number of sensors and
data size), accuracy and latency for real-time implementation. Based
on such analysis, in Sect.~\ref{sec:Protective-human--robot-distance}
we evaluate the robustness of the system in terms of the protective 
human--robot separation distance~$d_{p}$. Protective distance is 
underpinned by accuracy and latency, that are both quantified for 
the selected HRC functions. Concluding remarks and open problems are 
highlighted in Sect.~\ref{sec:Conclusions}.

\section{Multisensor data fusion platform: feature extraction, fusion and analytics }

\label{sec:Multisensory-edge-cloud-platform}

MDF platforms~\cite{ExManuf_2020} provide accurate, 
reliable and timely measurements of a physical process by combining sensor data. 
Sensors provide measurement data to the sensor fusion level where a fusion algorithm is used to extract and process information. 
MDF system architectures can be centralized, de-centralized, distributed or hierarchical~\cite{Sensors_2019}. 
The platform adopted here is depicted in Fig.~\ref{fig-edge-cloud}: it consists of distributed edge units each assigned to monitor an individual 
workspace area around a robot (\emph{i.e.}, the robotic cell) and a cloud 
server in charge of analytics tasks. The edge unit organizes the processing of data obtained from the corresponding sensors into~$N$ \emph{data
processing pipelines}, each managed by a dedicated server (\emph{micro-edge}).
Each data processing pipeline consists of a group of~$M$ sensors
whose raw data require similar pre-processing stages, including data
abstraction from raw radio signals and de-noising (\emph{i.e.}, background
subtraction). Considering~$M$ sensors and~$N$ pipelines, the raw
data~$\mathbf{X}_{k,i}(t)$ at time~$t$ obtained from sensor~$k=1,...,M$
and pipeline~$i=1,....,N$ follows sensor-specific pre-processing
stages represented by the operator~$f_{i}(\cdot|\phi_{i})$
\begin{equation}
\widetilde{\mathbf{X}}_{k,i}(t)=f_{i}\left(\mathbf{X}_{k,i}|\phi_{i}\right)\label{eq:intro}
\end{equation}
that uses pipeline-dependent parameters~$\phi_{i}$. 
Pre-processing is described in Sect.~\ref{architect} based on the chosen devices and sensing hardware. 
The edge units are in charge of extracting \emph{features} using the de-noised~$\widetilde{\mathbf{X}}_{k,i}(t)$ time series as inputs. 
The features, described in Sect.~\ref{subsec:Edge-node:-data}, are statistical, low dimensional, representations of~$\widetilde{\mathbf{X}}_{k,i}(t)$. Features have a uniform representations across different pipelines. 
Therefore, they can be easily combined, or fused, and jointly analyzed. 
The selection
of features to be fused depends on the implemented HRC function and it is
managed by the cloud server, through the data controller. Machine
learning (ML) tools for analytics on input features (classification)
are described in Sect.~\ref{subsec:Cloud-unit-and}.

\subsection{Edge node: data distribution, feature extraction and fusion}

\label{subsec:Edge-node:-data}
\begin{figure*}
	\centering
	\includegraphics[width=1\textwidth,trim={0 1cm 0 .4cm},clip]{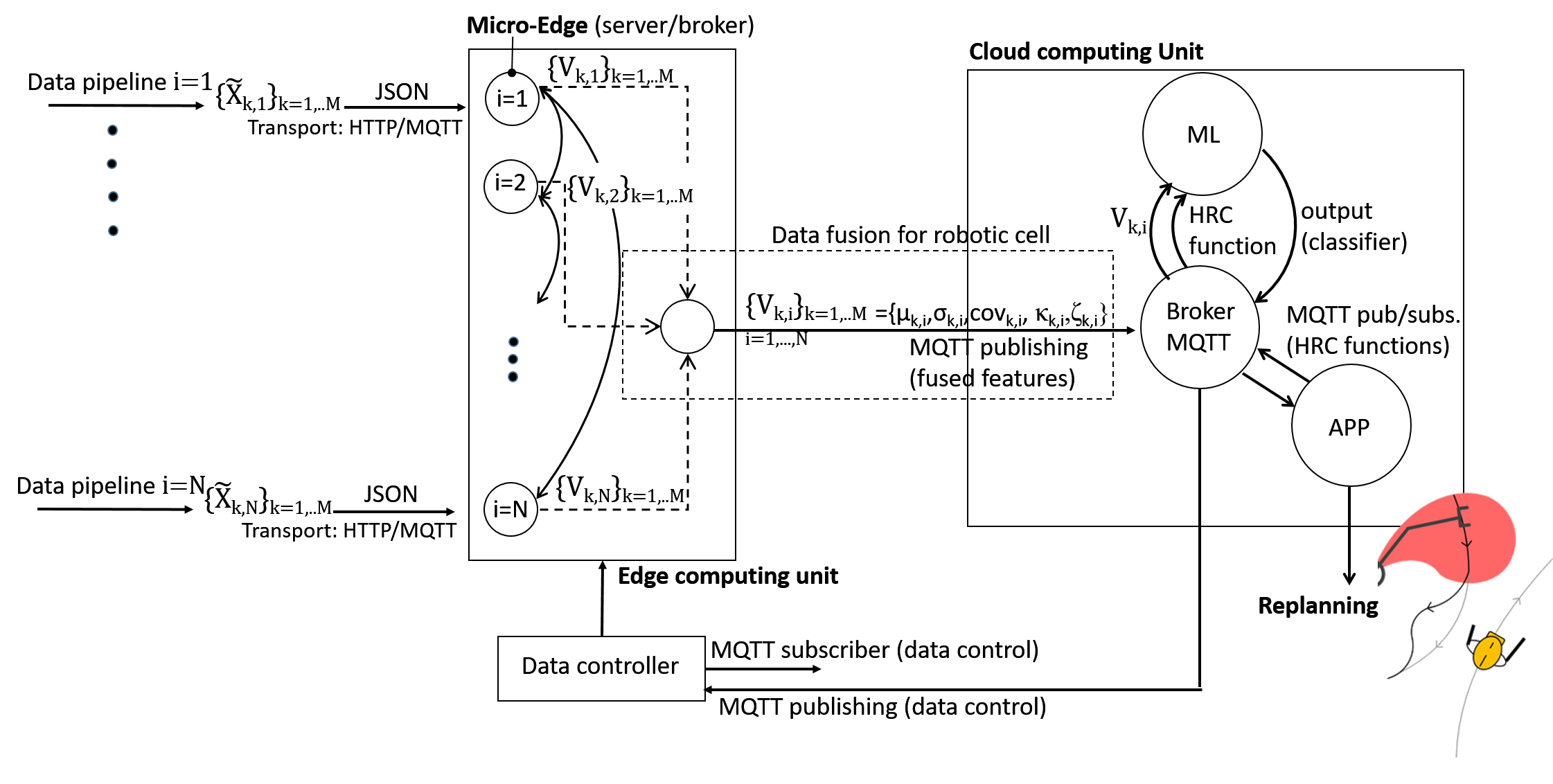} 
	\caption{Software architecture of the edge/cloud MDF platform: feature computing, 
		data fusion, distribution and transport layers.}
	\label{fig-edge-cloud-1} 
\end{figure*}

We focus here on data distribution at the edge of the MDF platform, and from the edge to the cloud unit as well. 
As shown in Fig.~\ref{fig-edge-cloud-1}, the edge node contains~$N$ micro-edges.
They act as dedicated servers/brokers for the corresponding data pipeline carrying de-noised data~$\mathbf{\widetilde{X}}_{k,i}(t)$. 
Considering telemetry collection on each pipeline, both the de-noised data~$\mathbf{\widetilde{X}}_{k,i}(t)$ and the corresponding time-stamp ($t$) information are represented by JavaScript Object Notation (JSON) structures. 
Sensor data~$\mathbf{\widetilde{X}}_{k,i}(t)$ is sent by the individual sensors through a wireless network supporting
both WiFi and IEEE 802.15.4 2.4 GHz standards (see Sect.~\ref{Experiment}
for further details). For transport level, both RESTful Web Services
and Message Queuing Telemetry Transport (MQTT) can be adopted inside
each micro-edge. Therefore, when using HTTP transport (RESTful) the micro-edge acts as a dedicated server exposing resources.
On the contrary, during MQTT transport, it acts as broker, accepting subscriptions from the specific pipeline for telemetry publishing.
Finally, the micro-edges also provide computing, storage and caching
functions~\cite{wot-sw} for the corresponding pipeline.

Each~$i$-th micro-edge is in charge of mid-level data analytics
for the corresponding pipeline. It processes the corresponding
de-noised data-sets~$\left\{ \mathbf{\widetilde{X}}_{1,i}(t),...,\mathbf{\widetilde{X}}_{M,i}(t)\right\}~$
to obtain the high level \emph{features}~$\mathbf{V}_{i}=\left\{ \mathrm{V}_{1,i},...,\mathrm{V}_{M,i}\right\}$
for the considered pipeline. Features are obtained individually at
the edge node; however, the cloud unit can supervise such stage through
a data controller feedback (Fig.~\ref{fig-edge-cloud-1}) that is used
to select the features to be processed based on the specific HRC function
as shown in Sect.~\ref{Experiment}. In particular, considering the data~$\mathbf{\widetilde{X}}_{k,i}(t)$
from pipeline~$i$ and sensor~$k$, four distinctive statistical representations
$\mathrm{V}_{k,i}=\left\{ \mu_{k,i},\sigma_{k,i},{\zeta}_{k,i},{\kappa}_{k,i}\right\}~$
are used as features where 
\begin{equation}
\begin{array}{c}
\mu_{k,i}=\mathbb{E}_{t}\left[\mathbf{\widetilde{\mathrm{X}}}_{k,i}(t)\right]\;\\
\sigma_{k,i}=\sqrt{\mathbb{E}_{t}\left[\left(\mathbf{\widetilde{\mathrm{X}}}_{k,i}(t)-\mu_{k,i}\right)^{2}\right]}\;\\
{\zeta}_{k,i}=\mathbb{E}_{t}\left[\left(\frac{\mathbf{\widetilde{\mathrm{X}}}_{k,i}(t)-\mu_{k,i}}{\sigma_{k,i}}\right)^{3}\right]\;\\
{\kappa}_{k,i}=\mathbb{E}_{t}\left[\left(\frac{\mathbf{\widetilde{\mathrm{X}}}_{k,i}(t)-\mu_{k,i}}{\sigma_{k,i}}\right)^{4}\right]\;.
\end{array}\label{eq:features}
\end{equation}
$\mu_{k,i}$ and~$\sigma_{k,i}$ are the mean and the standard
deviation, while~${\zeta}_{k,i}$ and~${\kappa}_{k,i}$ track the
skewness (\emph{i.e.}, third moment) and the kurtosis (\emph{i.e.}, fourth
moment) coefficients evaluated over consecutive time samples. As shown
in Fig.~\ref{fig-edge-cloud-1}, output features are again serialized
by the edge node using JSON representation and sent to the cloud using
MQTT as transport layer.

\subsection{Cloud unit: feature analytic tools and HRC functions}

\label{subsec:Cloud-unit-and}

The cloud unit processes the features (\ref{eq:features}) published
by the edges to infer a hidden process. Considering the three HRC
functions introduced in Sect.~\ref{subsec:Contributions}, the corresponding
hidden processes are: \emph{i}) the number of subjects in the workspace
(worker counting); \emph{ii}) their positions inside the operating
space (occupancy), and \emph{iii}) the human-robot distance~$d$ (co-presence
monitoring). The cloud uses a ML model to process a selected subset
of features. The ML model and the feature subset are optimized
using a supervised approach that is based on training data collected
inside the HRC environment. In particular, the Long-Short Term Memory
(LSTM) and the Convolutional Neural Network (CNN) models are chosen
for real-time manipulation of heterogeneous feature streams. Opportunities
and limitations of both models are considered in Sect.~\ref{Experiment}.

With respect to end-user interfacing, third party applications can
subscribe to the cloud, via MQTT messaging, for a specific HRC function
of interest, and wait for real-time results pushed back by the ML
unit, through the cloud service broker. The cloud unit thus provides
an open layer between end-user applications and the underlying feature
manipulation resources.

\begin{figure*}
	\includegraphics[width=1\textwidth]{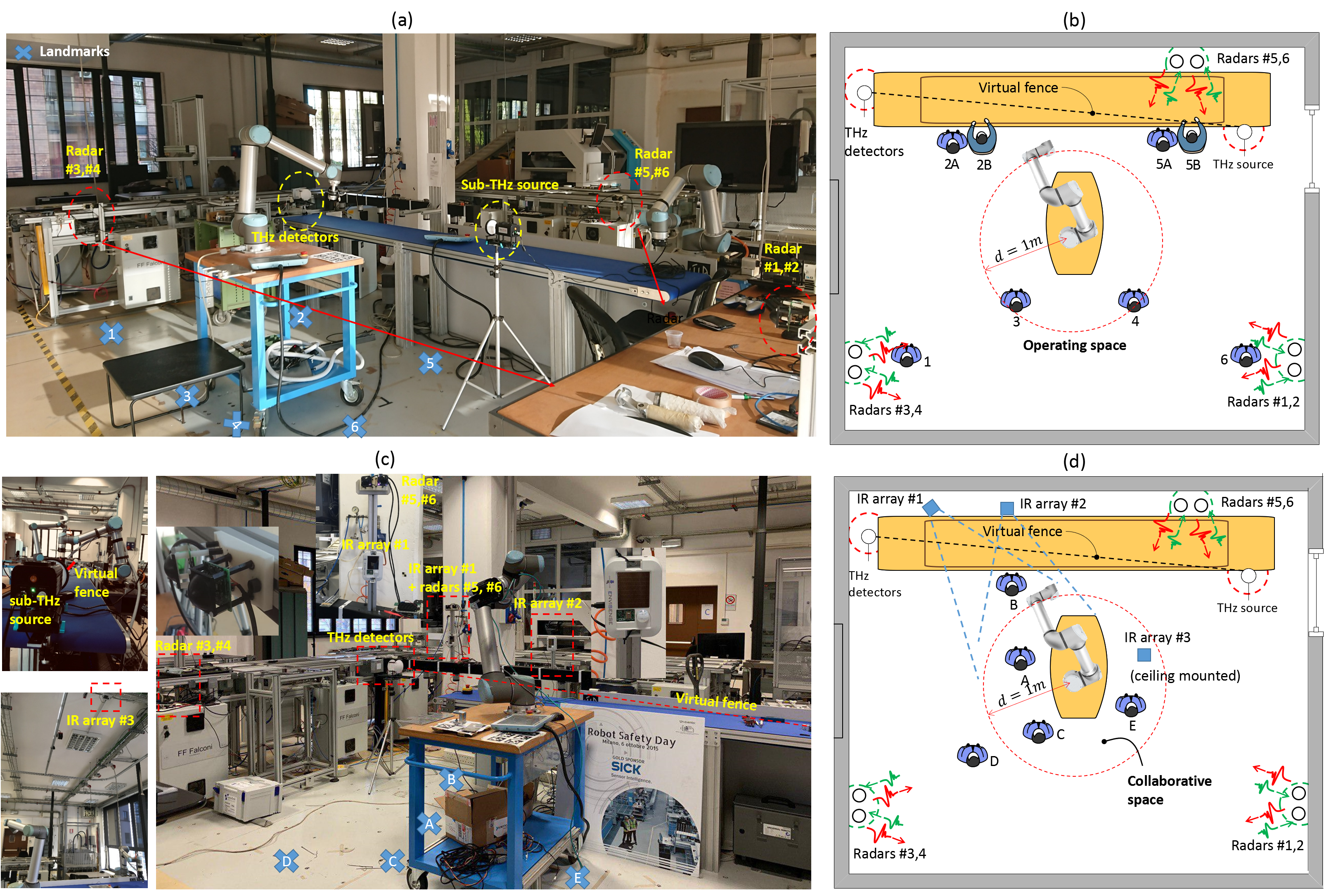} 
	\caption{Industrial pilot plant, cobot, and sensor deployment: a) robotic cell
		environment with sub-THz radar and detector deployment; b) landmark
		positions for worker motion detection ($d>1$~m); c) IR arrays and
		THz sensors deployment; d) landmark positions for worker-robot co-presence
		monitoring ($d<1$ m). }
	\label{fig-layout} 
\end{figure*}

\section{Opportunistic radio sensing inside the collaborative space}

\label{architect}

The robotic cell, depicted in Fig.~\ref{fig-layout}, consists of one 
manipulator performing a pick and place activity in proximity of an 
assembly line. The MDF platform adopts~$N=4$
pipelines and a wide range of IoT passive radio sensing devices operating
in the sub-THz ($100-122$ GHz), microwave ($2.4-5$ GHz), and long-wavelength
infrared (LWIR) ($8-13$~$\mu$m) bands from which it is possible to infer
the EM radiation as perturbed by the presence of the worker and the robot, 
or the thermal IR radiation as emitted by the human subject being measured, 
respectively. In this section, we describe the IoT sensors deployed, the 
signal model and the data pre-processing stages (equation~\ref{eq:intro}) implemented 
by the individual pipelines.

\subsection{Sub-THz radars and detectors}

\label{sensor} The sub-THz radiation refers to the frequency band
between~$0.1$ THz and~$1$ THz. In this band, the propagation is
extremely sensitive to environmental changes and presence of 
chemical/biological agents~\cite{tera_2018} typically affect both EM spreading and absorption losses. 
Beside sensing, THz band communication~\cite{AKYILDIZ201416} is also expected to be widely applied in industrial applications~\cite{tera_magazine}.
Compared to passive sensing technologies in the microwave~$1-50$
GHz band, the adoption of sub-THz signals is expected to provide precise
imaging in the short range, as well as reduced multipath effects,
to further increase the accuracy of human-scale sensing in exchange
of a more limited coverage area. The MDF platform combines two
sub-THz sensor systems managed by two distinct pipelines:
\emph{i}) a network of Frequency Modulated Continuous Wave
(FMCW) radars working at 122 GHz~\cite{122GHz_radar} that are deployed
for the passive localization of workers in the robotic cell, and \emph{ii})
a 2D array of co-located THz detectors~\cite{sub-TH_2019} monitoring
the intensity of the radiation emitted by a sub-THz generator ($100$
GHz) in the surroundings of the robot.

\begin{figure*}
	\includegraphics[width=1\textwidth]{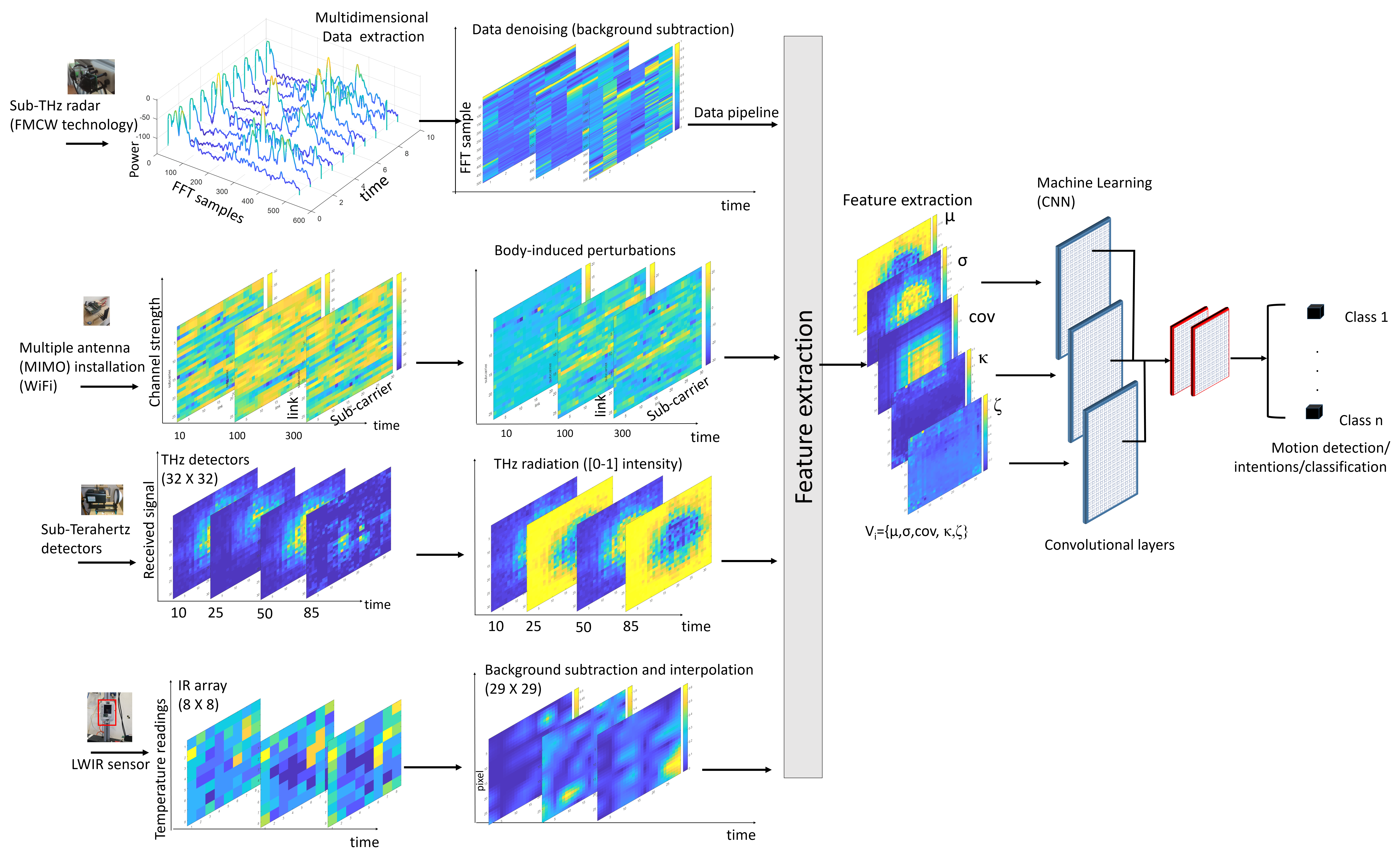} \caption{MDF platform and data fusion examples. From left to right:
		raw data and feature extraction from~$4$ data pipelines (sub-THz
		radars at~$122$ GHz - FMCW technology, WiFi, sub-THz detectors at
		$100$ GHz, and LWIR arrays), feature processing and classification.}
	\label{figureTeaser} 
\end{figure*}

\textbf{FMCW radars} are deployed along the perimeter of the robotic
cell and process the received beat signals corresponding to echoes
from a target/obstacle that are mixed with the transmitted signal.
The transmitted signal is a sweep frequency modulated (FM) continuous
wave (CW) with bandwidth up to~$6$ GHz and ramp (pulse) duration
of~$T=1$ ms. The carrier frequency is set in the nominal~$122$ GHz
ISM (Industrial Scientific and Medical) band. Beat signals are first converted
in the frequency domain by~$512$-point Fast Fourier Transform (FFT) and
averaged over~$8$ consecutive frames. Fig.~\ref{figureTeaser} on
top shows selected FFT images from~$M=6$ radars versus time, collected
in an empty and an occupied robotic cell. FFT images are clearly sensitive
to the presence of the worker. Assuming that radars are processed
by pipeline~$i=1$, for radar~$k\in\left[1,...,M\right]$ and time
$t$, the corresponding~$N_{FFT}=512$ FFT samples 
\begin{equation}
\mathbf{X}_{k,1}(t)=[x_{k,t}(1),...,x_{k,t}(N_{FFT})]
\end{equation}
undergo pre-processing steps, namely~$\widetilde{\mathbf{X}}_{k,1}(t)=f_{1}\left(\mathbf{X}_{k,1}|\phi_{1}=\left[\mathbf{C}_{k},\mathbf{\overline{X}}_{k}(\emptyset)\right]\right)$, for de-noising and background~$\mathbf{\overline{X}}_{k}(\emptyset)$
subtraction~\cite{Short-range FMCW} as 
\begin{equation}
\widetilde{\mathbf{X}}_{k,1}(t)=\mathbf{C}_{k}^{-\frac{1}{2}}\left[\mathbf{X}_{k,1}(t)-\mathbf{\overline{X}}_{k}(\emptyset)\right],\label{eq:back}
\end{equation}
where~$\mathbf{\overline{X}}_{k}(\emptyset)=\mathbb{E}_{t}[\mathbf{X}_{k,1}(t\,|\,\emptyset)]$
is the time average FFT of the beat signal~$\mathbf{X}_{k,1}(t\,|\,\emptyset)$
observed in the \emph{empty} robotic cell~$(\emptyset)$, namely the
background, while~$\mathbf{C_{\mathit{k}}}$ is the covariance matrix
$\mathbf{C_{\mathit{k}}}=\mathbb{E}_{t}\left[\left(\mathbf{X}_{k,1}(t\,|\,\emptyset)-\mathbf{\overline{X}}_{k}(\emptyset)\right)\,\left(\mathbf{X}_{k,1}(t\,|\,\emptyset)-\mathbf{\overline{X}}_{k}(\emptyset)\right)^{\mathrm{H}}\right]$.
Covariance takes into account any environmental change (\emph{i.e.},
due to concurrent robot movements) that might alter the backscattered
wavefield, yet without constituting any alert for worker safety. The
de-noised FFT time series~$\widetilde{\mathbf{X}}_{k,1}(t)$ are processed
by the edge unit to track the relative distance of the worker from
the robot, from which body movements as well as HR separation distance
($d$) can be inferred\footnote{we assume the manipulator (robot) position 
as known, analysis of human and robot position uncertainty is considered in Sect.~\ref{sec:Protective-human--robot-distance}}. Combining/fusing selected radar output features 
with other RF sensors enable more advanced worker perception modalities 
as discussed in Sect.~\ref{Experiment}.

\textbf{Sub-THz detectors} capture the sub-THz radiation originated
from a radio source (about~$80$ mW at carrier frequency~$100$ GHz)
generated by an IMPATT (Impact ionization Avalanche Transit-Time)
diode~\cite{impatt2007}. The~$M$ detectors ($M=1024$) are arranged
over a 2D array of~$(32\times32)$ elements forming a sub-THz camera
with rectangular shape. Each sub-THz detector, developed by~\cite{tera_2015},
is sensitive in the~$0.05-0.7$ THz band with noise-equivalent power
of 1 nW/$\sqrt{Hz}$. The 2D array captures the presence of the worker
moving in the surrounding of the line-of-sight (LOS) path connecting
the sub-THz source and the camera. Considering the pipeline~$i=2$,
$\mathrm{X}_{k,2}(t)$ is a measure of the~$[0-1]$ normalized intensity,
or strength, of the radiated sub-THz field observed by the detector
$k\in\left[1,...,M\right]$ of the array at time~$t$. 
Likewise, for FMCW radars, Fig.~\ref{figureTeaser} shows an example of the normalized
signal intensity measured by the 2D array. The measured intensity
is proportional to the electric field power of the received electromagnetic
waves: any change of the received signal intensity corresponds to
different worker body parts obstructing the radio link (\emph{e.g.},
torso, arms and hands). In general, the multi-ray propagation between
the fixed sub-THz source and the~$k$-th detector consists of line-of-sight
(LOS), reflected, scattered and diffracted rays~\cite{sub-TH_2019}.
When no obstacles are present near the LOS path and the sub-THz setup
is directional (\emph{e.g.}, using dielectric lenses as in the experimental
setup), the propagation is mostly due to the main LOS ray that depends
only on the free-space loss and the absorption loss. Therefore, the
radiation intensity is modelled as

\begin{equation}
\mathrm{X}_{k,2}(t)=\mathrm{S}_{k}(t)+\mathrm{W}_{k}(\emptyset),\label{eq3-1}
\end{equation}
where~$\mathrm{S}_{k}(t)$ corresponds to the body-induced signature
and~$\mathrm{W}_{k}(\emptyset)\sim\mathcal{N}(b_{k},\sigma_{k}^{2}$)
models the background radiation observed in the empty robotic cell.
Each background component~$\mathrm{W}_{k}$ is modeled here as an
independent Gaussian term with average intensity~$b_{k}$ and deviation
$\sigma_{k}^{2}$. The average component~$b_{k}$ is a function of
LOS terms: absorption and free-space loss~\cite{sub-TH_2019}. Reflection,
scattering and diffraction effects due to a worker or the robot near
the LOS path introduce time-varying changes of radiation intensity
modeled by deviation~$\sigma_{k}^{2}$. Post-processed data~$\mathbf{\widetilde{\mathrm{X}}}_{k,2}(t)$
consists of signatures that are estimated from the input radiation
measurements~$\mathrm{X}_{k,2}(t)$ through subtraction and de-noising,
$\widetilde{\mathrm{X}}_{k,2}(t)=f_{2}\left(\mathrm{X}_{k,2}|\phi_{2}=\left[b_{k},\sigma_{k}\right]\right)$,

\begin{equation}
\mathbf{\widetilde{\mathrm{X}}}_{k,2}(t)=\sigma_{k}^{-1}(\mathrm{X}_{k,2}(t)-b_{k}).\label{eq5}
\end{equation}
In Sect.~\ref{Experiment}, we exploit the sub-THz radio technology to implement a virtual safety fence and monitor the worker presence close by the assembly line.

\subsection{IR array sensors and body-induced thermal signatures}

\label{irarray} The use of thermal sensors for human body sensing~\cite{thermal2}
is becoming attractive in many IoT-relevant scenarios, such as smart
spaces, assisted living and industrial automation~\cite{thermal3}.
Thermal vision and related computing tools enable the possibility
of analyzing body induced thermal signatures for detection of body
motions, fall detection, as well as discriminating those signatures
from the environment. The experimental validation scenario of 
Sect.~\ref{Experiment} exploits~$M=3$ sensors. Each sensor consists
of an array of~$N_{TP}=64$ thermopile detectors that are sensible
in the~$8-13$~$\mu$m LWIR infrared band with a noise equivalent
temperature difference of~$\pm0.08$ °C @ 1 Hz at room temperature.
Each sensor acquires thermal IR images organized as 2D frames of~$8\times8$
pixels~\cite{thermal4}.

Thermopile detectors independently measure the captured IR radiation
and convert it to temperature readings that are the inputs of the
data pipeline~$i=3$ (in Fig.~\ref{fig-edge-cloud}) 
\begin{equation}
\mathbf{X}_{k,3}(t)=[x_{k,t}(1),...,x_{k,t}(N_{TP})].
\end{equation}
Pre-processing of temperature readings is carried out to obtain the
body-induced thermal signatures~$\widetilde{\mathbf{X}}_{k,3}(t)=f_{3}\left(\mathbf{X}_{k,3}|\phi_{3}=\left[\mathbf{W}_{k}(\emptyset)\right]\right)$
that measure the temperature increase as induced by body movements
\begin{equation}
\widetilde{\mathbf{X}}_{k,3}(t)=\mathbf{X}_{k,3}(t)-\mathbf{W}_{k}(\emptyset)
\end{equation}
where now~$\mathbf{W}_{k}(\emptyset)=[x_{k,t}(1),...,x_{k,t}(N_{TP})]$
conveys information about stationary heat-sources (\emph{i.e.}, robots,
other machinery) that are not caused by body movements but characterize
the background temperature of the empty space~$\emptyset$. Computer
vision methods~\cite{thermal2} or statistical approaches~\cite{thermal3}
can be adopted to track body positions/motions. In Sect.~\ref{Experiment},
we verify that fusion of such information with features obtained from
THz radars is effective in improving worker motion discrimination
in proximity with the robotic manipulator.

\subsection{Multi-antenna installations for directional sensing and worker counting}

\label{subsec:counting}

Increasingly, multi-antenna systems are installed in WiFi and cellular
settings. Such antenna-diversity techniques can be exploited in I4.0
scenarios, for instance, for simultaneous multi-target tracking, for
worker counting, or for recognition of complex human contexts or activities.
We assume here a scenario with a dedicated transmitter and a multi-antenna
receiver in the direct LOS. Such installation might span a workplace
scenario with multiple work locations. At the receiver side, various
phase offsets of the RF-chains correspond to different steering angles
of the recognition beams. It is thus possible to enable directional
perception of the environment from multi-antenna omnidirectional devices.
This angular information establishes evidence on the direction of
body movement with respect to the antenna. At the same time, multiple beam
analysis can further grant simultaneous multi-target recognition and
tracking. This enables, for instance, the simultaneous tracking and
monitoring of multiple body parts, as well as the counting of multiple
subjects.

In what follows, we assume a WiFi installation featuring a multi-antenna receiver equipped with an antenna array with~$M$ elements that
is connected to pipeline~$i=4$. The received signal at~$k$-th antenna
of the array ($k=1,...,M$), namely the channel state information
(CSI)~$\mathrm{X}_{k,4}(t)$, is composed~\cite{palipana2019extracting}
by the direct LOS signal~$\mathrm{L}_{k}(t)$, the signal~$\mathrm{i}_{k}(t)$
reflected by the target, as well as noise~$\mathrm{n}_{k}(t)$: 
\begin{equation}
\mathrm{X}_{k,4}(t)=\mathrm{L}_{k}(t)+\mathrm{i}_{k}(t)+\mathrm{n}_{k}(t).
\end{equation}
It is possible to amplify and isolate the desired signal
$\mathrm{i}_{k}(t)$ by applying proper beamforming weights~$\mathbf{w:=}\left\{ w_{k}\right\} _{k=1}^{M}$
and by subsequently subtracting the estimated LOS signal~$\widetilde{\mathrm{L}}_{k}(t)$~\cite{sameera2019dfhs}. The de-noised sequence~$\widetilde{\mathrm{X}}_{k,4}(t)=f_{4}\left(\mathrm{X}_{k,4}|\phi_{4}=\left[\mathbf{w},\widetilde{\mathrm{L}}_{k}(t)\right]\right)$
serves as an estimate of~$\mathrm{i}_{k}(t)$
and can be further processed to detect movements or activities. Repeated
correction of the beamformer weights~$\mathbf{w}$ achieves mobile
target tracking and activity recognition~\cite{palipana2019extracting}.

For a high recognition accuracy, it is beneficial to establish a small
beam width. The beam width is naturally dictated by the count ($M$)
of antennas at the receiver. In the presence of multiple targets (\emph{e.g.},
multiple subjects or multiple moving body parts), isolation of distinct
array responses for various directions yields separate information
about these multiple targets and thus allows multi-subject or multi-target
tracking~\cite{palipana2020PerCom}. It has been shown in~\cite{palipana2020PerCom},
that proper clustering of this spatial streams can be used for crowd
counting applications. In Sect.~\ref{Experiment}, the worker counting
system is implemented to monitor a passage room where workers move \emph{e.g.}
before entering the shared workspace.

\section{Experimental validation and case studies}

\label{Experiment}

The proposed IoT MDF platform is validated through experimental
activities inside a representative robotic cell~\cite{Dimostratore}. 
In particular, we focused on~$3$ scenarios, each targeting a different 
HRC function: 
\begin{itemize}
\item \emph{Worker counting}: In Sect.~\ref{subsec:countingCase}, we tackle 
the problem of counting the number of workers present in a monitored passage 
room of size 5.5 m x 4 m  by exploiting directional information on signal 
perturbations, obtained from multi-antenna instrumentation 
(pipeline~$i=4$, Sect.~\ref{subsec:counting}). The system employs 
training-free techniques and thus it is robust to changing environments. 
\item \emph{Worker motion detection}: Sect.~\ref{subsec:motion} addresses
the problem of worker motion detection inside the \emph{operating
space} (approx.~$d>1$~m from the robot) of the robotic cell and for
different positions of the worker. The corresponding layout is detailed
in Fig.~\ref{fig-layout}.a,b. The proposed MDF platform applies feature
fusion by using sub-THz radars (pipeline~$i=1$) and sub-THz detectors ($i=2$). The scenario is effective in many SSM tasks as shown in Sect.~\ref{subsec:hrc_cooperation}. 
\item \emph{Worker-Robot co-presence monitoring}: Sect.~\ref{subsec:copresence}
addresses the critical case of human--robot workflow monitoring inside
the \emph{shared space} and during a collaborative task that
requires the worker and the robotic manipulator to cooperate at close
distance ($d<1$ m). The corresponding layout is detailed in Fig.~\ref{fig-layout}.c
and~\ref{fig-layout}.d. For this case, the platform combines both
sub-THz radar sensors ($i=1,2$) and IR array devices ($i=3$) that are re-used here opportunistically to track the worker motion intentions and occupancy (\emph{i.e.},
position) inside the working area. 
\end{itemize}
The collaborative setup is composed by a robotic manipulator from Universal Robot
UR10, with~$10$ kg payload mounted on a workdesk. Robot tasks are
pick-and-place activities: the robot picks objects from an inbound
conveyor and moves them to a outbound bin. In the meantime, the operator
picks the objects that require addition work, such as dissembling
or re-manufacturing.

As shown in Fig.~\ref{figureTeaser}, the IoT MDF platform
consists of~$N=4$ selectable pipelines described in Sect.~\ref{architect},
namely: \textit{i})~$6$ FMCW radars (Sect.~\ref{sensor}); \textit{ii})
a sub-THz source and camera (Sect.~\ref{irarray}) implementing a
virtual fence system; \textit{iii})~$3$ IR array sensors, with one
of them ceiling-mounted; and \textit{iv})~$3$ software-defined radio devices to test worker counting functions (Sect.~\ref{subsec:countingCase}).
Sensors continuously monitor the area by collecting raw data and forwarding
them to the corresponding micro-edge server. For data distribution,
FMCW radars and sub-THz cameras use a 2.4 GHz WiFi link, while IR
sensors send the data to an intermediate access point (AP) using an
IEEE 802.15.4e implementation~\cite{IoT_2018}, the AP forwards the
aggregated data to the edge via 2.4GHz WiFi. The edge node extracts 
and sends the features to the cloud for classification
(Sect.~\ref{sec:Multisensory-edge-cloud-platform}). 
Features are transferred through MQTT messages in real-time using a 
wired transport network (Ethernet), but optimized for safety~\cite{IoT2016}. 

A data controller handles the orchestration of features: it can
therefore decide how many pipelines should be combined inside the edge 
depending on the required HRC function. For example, occupancy detection 
inside the operating space does not require high accuracy, considering 
that the worker is operating at safe distance from the robot. It could
be obtained using only a couple of data pipelines to minimize latency. 
On the contrary, co-presence monitoring requires higher accuracy, therefore features must be selected considering the latency-accuracy trade-off and the protective distance~$d_{p}$ evaluation. In what follows, both non-cooperative sensing, that uses the individual pipelines separately, and cooperative sensing, that adopts the fusion of features, are discussed and compared using boht CNN and LSTM models.

\begin{figure}
\includegraphics[width=0.5\textwidth]{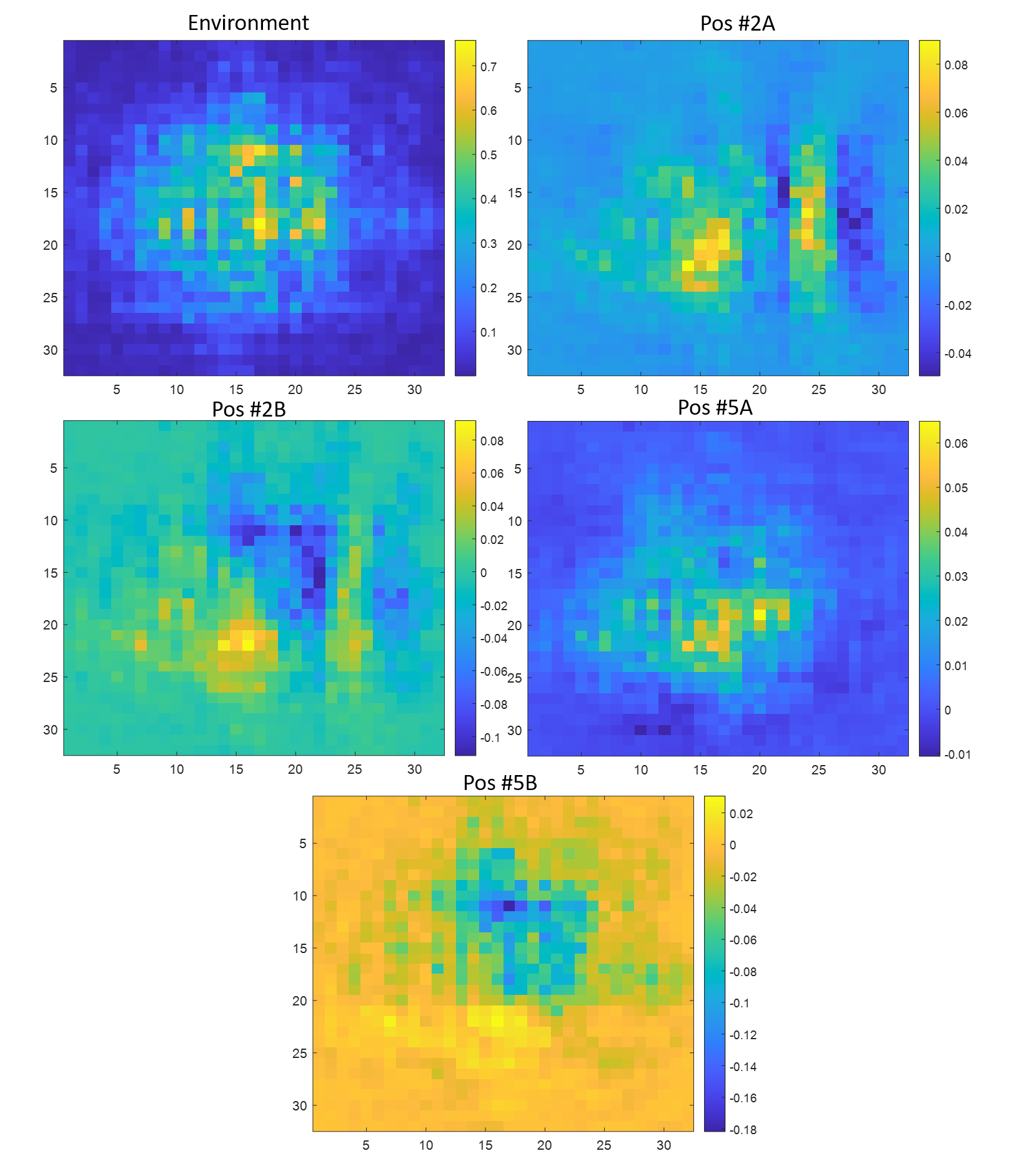} \caption{Image 
of human body movements captured by sub-THz detector array at
the predefined locations mentioned in layout in Fig.~\ref{fig-layout}. 
Position number at images corresponds to predefined locations A/B 
correspond to human body wo/w body movements, respectively.}
\label{figureTera} 
\end{figure}

\subsection{Worker counting}

\label{subsec:countingCase}

The first use case focuses on the worker counting problem using features
from pipeline~$i=4$ and obtained from a multi-antenna radio installation.
To this aim, in an indoor area of 5.5 m x 4.0 m, we installed a dedicated
transmitter (USRP X300 series with UBX-160 and SBX RF-daughterboard)
centered at one of the longer walls and a four-antenna receiver ($M=4$)
centered on the opposite wall. Another USRP is installed in the corner
of the area and used to synchronize with the Rx USRPs by
broadcasting a reference signal to calibrate each RX frame offset.
Clock coherence at each USRP Rxs has been achieved via a clock distribution
system, which provided a pulse-per-second and a 10 MHz reference signal
to discipline the local oscillators. This OFDM system operated at
3.42 GHz with a bandwidth of 15.32 MHz into which 5408 subframes (each
3082 samples long) are established over 52 usable subcarriers.

A total of 6 different people participated in the experiment and conducted
dedicated movements at up to 12 confined locations. Table~\ref{tableConfusion}
summarizes the results for 0-3 persons\footnote{The worker count is limited 
by the number of antennas utilized~\cite{palipana2020PerCom}}. We estimate 
the worker count within an error of~$1$ for up to~$4$
persons. Results were achieved in a training-free manner so that the
system is robust against frequent environmental changes or also to
re-location. In particular, this is useful to meet the high demands
of frequently and dynamically changing industrial workplace settings.
In particular, we detect workers via beam scanning in which the direction
of arrival for reflected and LOS signals create a distribution over
time. For environments with multiple workers present, the resulting
mixture of several Gaussian distributions is separated by a clustering
approach. We exploit joint approximate diagonalization of eigen-matrices
(JADE)~\cite{cardoso1993blind} for blind source separation of the
observed, only partially orthogonal streams. We derive a distance
matrix~$\mathbf{D}$ between the respective streams by applying constrained
derivative dynamic time warping (cDDTW)~\cite{keogh2001derivative}.
As clustering method, we have chosen Hierarchical agglomerative clustering
(HAC)~\cite{sharma2012comparison}, using the distance matrix~$\mathbf{D}$.

\begin{table}
\caption{Worker counting confusion matrix. Actual vs. estimated count (\% and
\#) in an indoor confined area of 5.6 m x 4.0 m.}
\label{tableConfusion} \resizebox{\columnwidth}{!}{ %
\begin{tabular}{cccccc}
\multicolumn{6}{c}{\textbf{Estimated worker count \% (\#)}}\tabularnewline
\multicolumn{1}{c|}{} & \multicolumn{1}{c}{\textbf{0}} & \multicolumn{1}{c}{\textbf{1}} & \multicolumn{1}{c}{\textbf{2}} & \multicolumn{1}{c}{\textbf{3}} & \textbf{4}\tabularnewline
\hline 
\multicolumn{1}{c}{\textbf{0}} & \cellcolor{100!100}{100\% (12)}  &  &  &  & \tabularnewline
\multicolumn{1}{c}{\textbf{1}} &  & \cellcolor{100!50}{89\% (16)}  & \cellcolor{47!40}{11\% (2)}  &  & \tabularnewline
\multicolumn{1}{c}{\textbf{2}} &  & \cellcolor{11!20}{3\% (1)}  & \cellcolor{89!47}{47\% (14)}  & \cellcolor{89!47}{47\% (14)}  & \cellcolor{11!20}{3\% (1)}\tabularnewline
\multicolumn{1}{c}{\textbf{3}} &  &  & \cellcolor{11!30}{5.5\% (1)}  & \cellcolor{100!50}{89\% (16)}  & \cellcolor{11!30}{5.5\% (1)} \tabularnewline
\end{tabular}} 
\end{table}

\begin{figure}
	\centering \includegraphics[width=1\columnwidth]{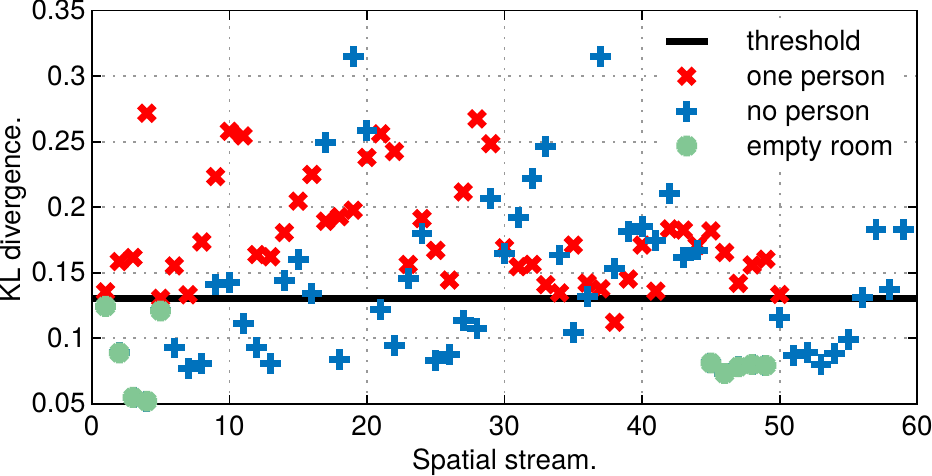} 
	\caption{Kullback-Leibner divergence of the detection of human absence in spatial
	streams.}
	\label{fig:Kullback} 
\end{figure}

The above approach is able to recognize the presence of workers in
distinct spatial streams. Since the amplitude fluctuation from neighboring 
streams is smaller than the one caused by a person in that spatial stream, 
noise and interference from neighboring locations
are small and hence the recognition accuracy improved. However, in
our case studies, noise and interference from neighboring locations
are not completely erased by the approach. Instead, some correlation
remains and it is particularly pronounced when the respective neighboring
cell is empty. In such case, false positives are likely and have to
be separately addressed by analyzing the correlation between spatial
streams with predicted worker presence. % If the correlation is high, a false positive is likely.

To show this, we computed the Kullback-Leibner divergence for pairs
of neighboring spatial streams where either: \textit{i}) no worker was
present in all spatial streams (namely \textit{empty room}), 
\textit{ii}) the stream was occupied by a worker (namely, \textit{one
person}), and \textit{iii}) the stream was not occupied but a neighboring
stream was (namely, \textit{no person}). Fig.~\ref{fig:Kullback} depicts
the Kullback-Leibner divergence for these samples.

%  We observe that, t
The empty room is well distinguished from the other cases, as most
%cases 
remain below the Kullback-Leibner threshold, and also the majority
of cases with \textit{one person} in the spatial stream are well above
the threshold. However, the cases with no person next to a stream
that is occupied by a person are not clearly distinguished and hence
are a potential source for false positives. To avoid this one-sided
error, we suggest to employ a correlation analysis over all spatial
streams. In fact, high correlation among neighboring streams indicates 
likely false positives.

\begin{figure}
	\centering \includegraphics[width=0.52\textwidth]{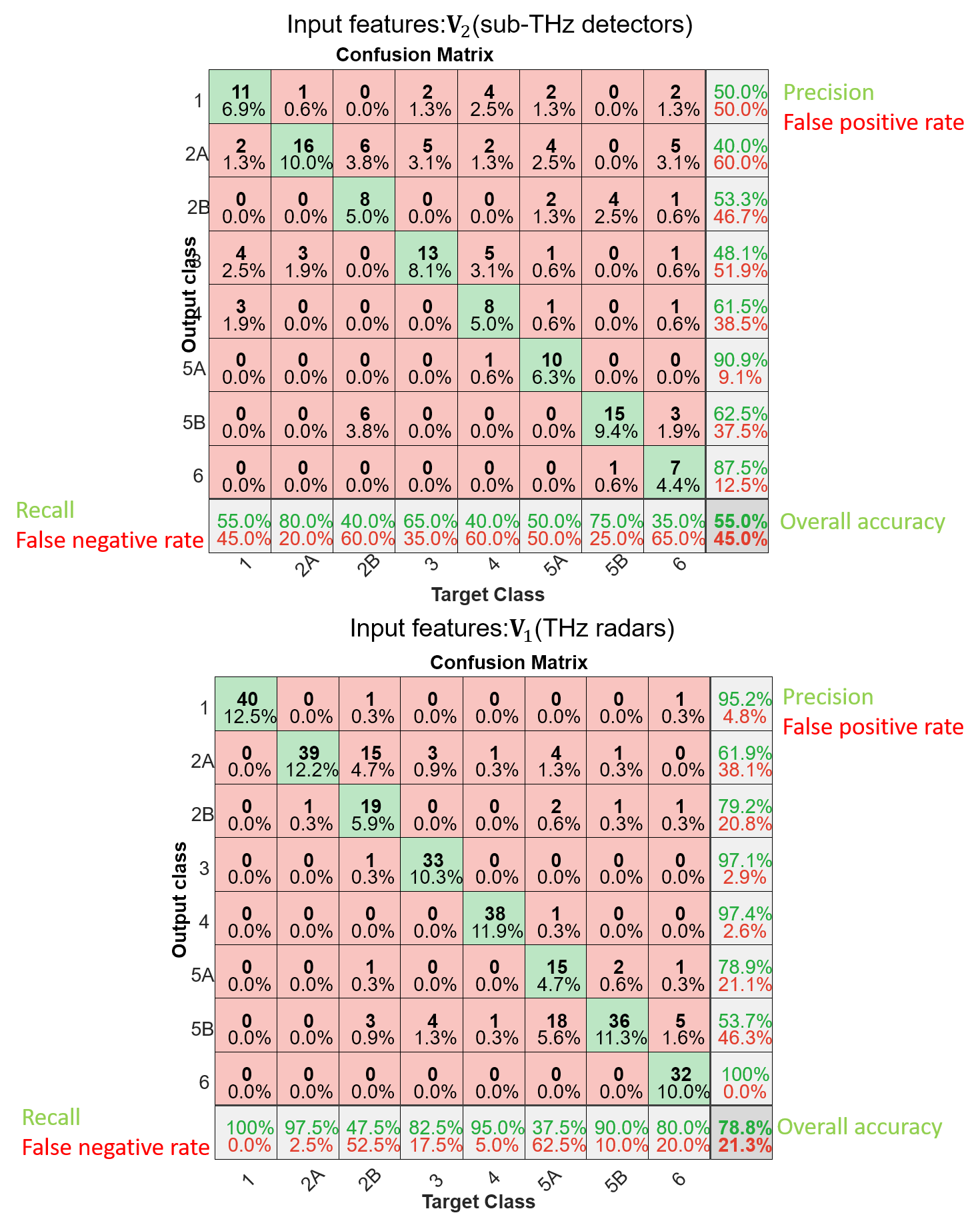}
	\caption{Confusion matrix results for LSTM networks for worker detection. Top:
		performances using only feature~$\mathrm{\mathbf{V}}_{2}$ from sub-THz
		detectors (pipeline~$i=2$) without data fusion; bottom: performances
		using only feature~$\mathrm{\mathbf{V}}_{1}$ from FMCW radars(pipeline
		$i=1$) without data fusion. The classes correspond to 8 activities
		of a worker at 6 predefined positions.}
	\label{ind_CM_I} 
\end{figure}

\begin{figure}
	\centering \includegraphics[width=0.52\textwidth]{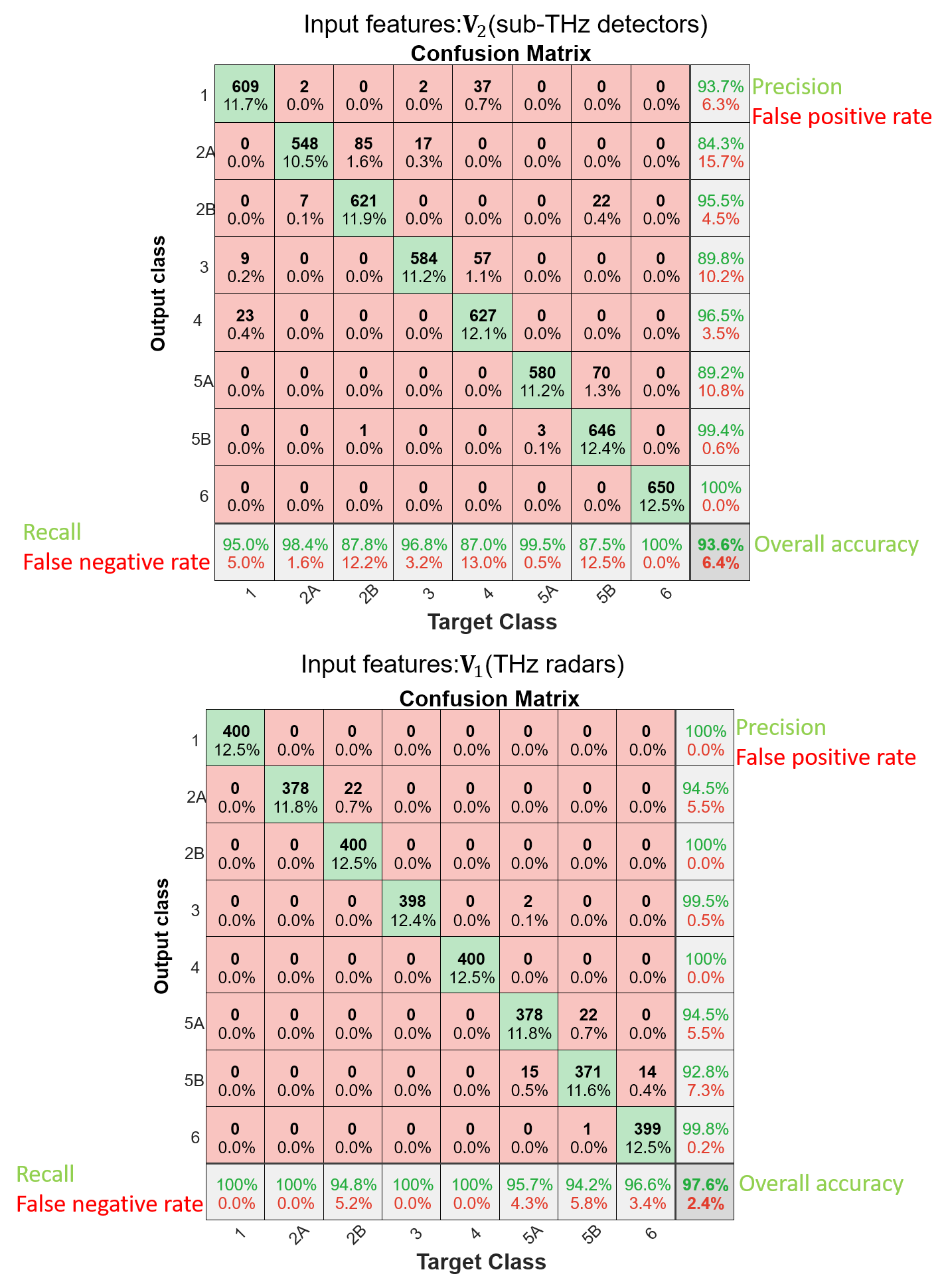}
	\caption{Confusion matrix results for CNN networks for worker detection. Top:
		performances using only feature~$\mathrm{\mathbf{V}}_{2}$ from sub-THz
		detectors (pipeline~$i=2$) without data fusion; bottom: performances
		using only feature~$\mathrm{\mathbf{V}}_{1}$ from FMCW radars (pipeline
		$i=1$) without data fusion. The classes correspond to the 8 activities
		of a worker at 6 predefined positions.}
	\label{ind_CM_II} 
\end{figure}

\begin{figure}
	\centering \includegraphics[width=0.52\textwidth]{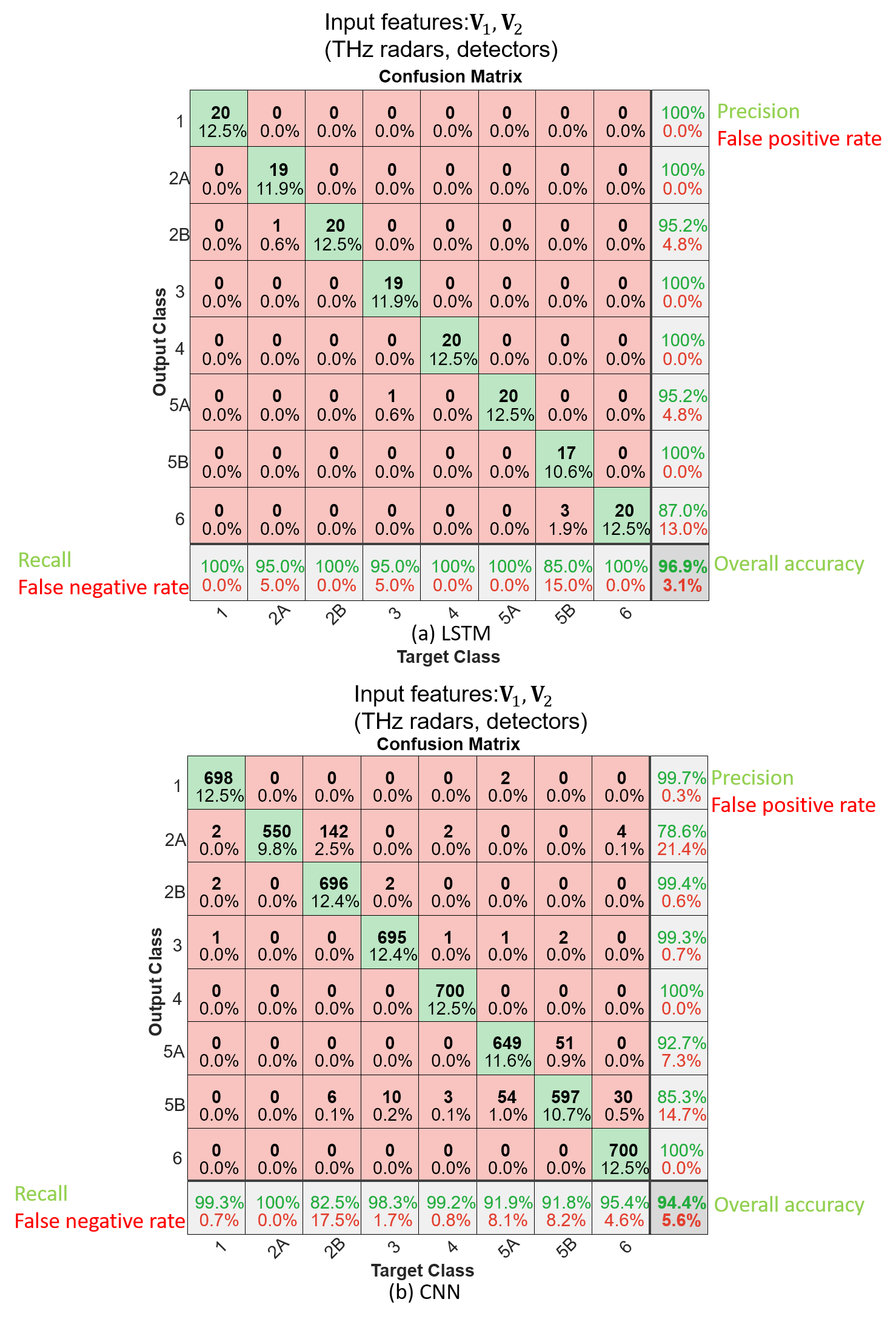}
	\caption{Confusion matrix results for worker detection using fused features
		$\mathrm{\mathbf{V}}_{1}$ and~$\mathrm{\mathbf{V}}_{2}$ from both
		FMCW radars (pipeline~$i=1$) and the sub-THz camera (pipeline~$i=2$).
		Top: LSTM performances; bottom: CNN performance. The classes correspond
		to the 8 activities of a worker at 6 predefined positions.}
	\label{sensorfusion_CM} 
\end{figure}

\subsection{Worker detection inside the operating space ($d>1$~$\mathrm{m}$)}

\label{subsec:motion}

The second scenario focuses on the robotic cell of Fig.~\ref{fig-layout}.a
and tests the performance of worker motion detection and localization
in the landmark positions~$1$ up to~$6$ inside the operating space of
the manipulator with distance~$d>1$~m from the robot. Since the worker, 
even when standing, is not rigid but moves around each landmark, the 
$6$ positions correspond to~$8$ different body postures:~$1,2\mathrm{A},2\mathrm{B},3,4,5\mathrm{A},5\mathrm{B},6$ as marked in the figure. 
For example, the worker at position number~$2$ implements two different postures:~$2\mathrm{A}$ relates to the worker standing
at position~$2$ and moving in the surroundings,~$2\mathrm{B}$ corresponds
to a worker performing simple pick-and-place activities in the surrounding
of the assembly line. Likewise, the same postures are used also in position~$5$.
In what follows, we analyze at first, the system performance considering the processing of the FMCW radar features~$\mathrm{\mathbf{V}}_{1}$ (pipeline~$i=1$) and the sub-THz detector ones~$\mathrm{\mathbf{V}}_{2}$ (pipeline~$i=2$) separately.
Next, sensor fusion is evaluated by merging the features~$\mathrm{\mathbf{V}}_{1}$,~$\mathrm{\mathbf{V}}_{2}$ and running ML on the cloud back-end.

Fig.~\ref{figureTeaser} on top shows an example of radar FFT samples
obtained from~$6$ FMCW sensors working in the~$122$ GHz band inside
the monitoring area. FFT samples are processed as in equation (\ref{eq:back}) from 
pipeline~$i=1$ inside the edge to extract the features~$\mathrm{\mathbf{V}}_{1}$ 
in (\ref{eq:features}). Features are then arranged and interpolated to 
fit with the required 2D ML model input size and sent to the ML stages 
running in the cloud.

Likewise, considering now pipeline~$i=2$, the radiation intensity 
obtained from the sub-THz camera is altered by the presence of the 
worker in the surroundings of the assembly line. This is visualized 
in Fig.~\ref{figureTera}, where we highlight the measured radiation 
obtained from the sub-Thz camera arranged as a 2D matrix of~$32\times32$ 
equivalent pixels in correspondence of the empty environment and when 
the worker is located at positions~$\mathrm{2A},2\mathrm{B},5\mathrm{A}$ and~$5\mathrm{B}$.
The sub-THz camera is reused opportunistically to detect and
discriminate worker motions in the surroundings of the assembly line.
A worker approaching the assembly line affects the received sub-THz 
radiation: the radiation intensity changes are more significant at 
positions~$2\mathrm{B}$ and~$5\mathrm{B}$, compared with postures 
at~$2\mathrm{A}$ and~$5\mathrm{A}$, due to the movements of the worker arms, in addition to the torso.

At the cloud back-end, LSTM and CNN tools are compared for position/activity
classification. In particular, the CNN model captures recurring structures 
from the input 2D features using spatially located convolutional filters. On 
the contrary, the LSTM network monitors long and short-term correlations over 
the feature samples, arranged similarly as for CNN. LSTM consists of an 
individual trainable layer of~$8$ units and~$8$ softmax outputs to track 
the corresponding worker positions. 
The CNN model uses 2D input features of size ($32\times32$) and consists of~$3$ trainable layers: it is detailed in Table~\ref{tableML} (for the case~$d>1$~m). 
Both non-cooperative and cooperative sensing cases use the same ML model. 
Note that, when fusing both features~$\mathrm{\mathbf{V}}_{1}$ and 
$\mathrm{\mathbf{V}}_{2}$, input size adaptations are required to fit with~$32\times32$ 2D dimensions, \emph{e.g.} by interpolation/decimation and reshaping. In addition, convolutional filters run independently for each feature/pipeline:
a flattening layer (see also Fig.~\ref{figureTeaser}) is used to 
collapse the convolutional filter output dimensions so that the 
outputs can be processed by the FC layer. In all cases,~$80$$\%$ 
of data is used for training and~$20$$\%$ for real-time classification. 

Figs.~\ref{ind_CM_I} and~\ref{ind_CM_II} show the confusion matrix (precision, recall, and overall accuracy) obtained with LSTM and CNN, respectively. 
Each tool uses individual features $\mathrm{\mathbf{V}}_{1}$ and~$\mathrm{\mathbf{V}}_{2}$ as inputs without implementing any data fusion (non-cooperative sensing).  
Fig.~\ref{ind_CM_I} on top shows the performance of the sub-THz camera 
system ($i=2$) while the bottom one shows 
the corresponding confusion matrix results obtained using the FMCW radars ($i=1$). 
Processing of features~$\mathrm{\mathbf{V}}_{1}$ gives 
better accuracy results, (\emph{i.e.},~$78.8$$\%$) with respect to~$\mathrm{\mathbf{V}}_{2}$ (\emph{i.e.},~$55$$\%$). In 
Fig.~\ref{ind_CM_II}, we now employed the CNN model by analyzing, again separately, 
the features~$\mathrm{\mathbf{V}}_{1}$ and~$\mathrm{\mathbf{V}}_{2}$.
We noticed that CNN outperforms LSTM since both features are more
spatially- than time-correlated.

In Fig.~\ref{sensorfusion_CM}, the detection system fuses both features~$\mathrm{\mathbf{V}}_{1}$ and~$\mathrm{\mathbf{V}}_{2}$ using data normalization techniques based on the min-max approach~\cite{handbook_reg}. 
Fig.~\ref{sensorfusion_CM} shows the confusion matrices for both LSTM and CNN. 
Results confirm the expected improvements of the cooperative approach that jointly operated on both~$\mathrm{\mathbf{V}}_{1}$  and~$\mathrm{\mathbf{V}}_{2}$ features. 
In fact, the accuracy increases to~$96.9$$\%$ and~$94.4$$\%$ for CNN and LSTM, respectively.

In what follows, the CNN model is chosen for protective distance~$d_{p}$ and quantitative evaluation of robustness (Sect.~\ref{sec:Protective-human--robot-distance}). 

\begin{figure}
	\includegraphics[width=0.5\textwidth]{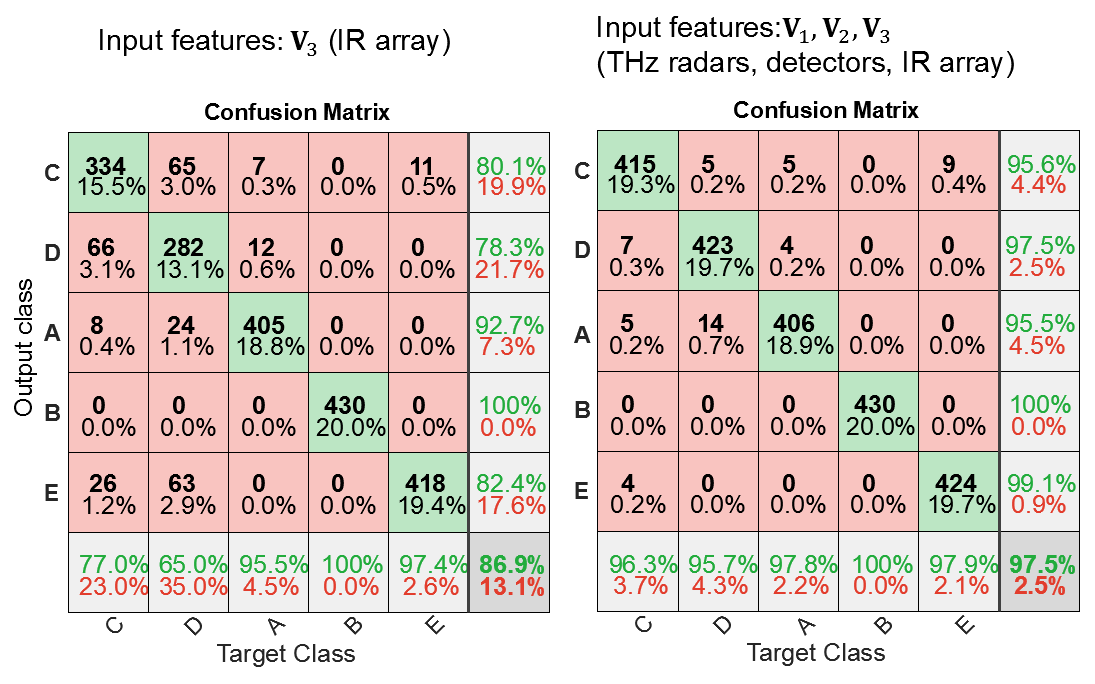} \caption{Confusion matrix results for worker-robot co-presence monitoring.
		Left: CNN performance without data fusion using only feature~$\mathbf{V}_{3}$
		from the IR array (pipeline~$i=3$). Right: CNN performance with data
		fusion using features~$\mathbf{V}_{1},\mathbf{V}_{2}$, and~$\mathbf{V}_{3}$
		from FMCW radars, the sub-THz camera, and the IR array (pipelines
		$i=1,2,3$), respectively.}
	\label{sensorfusion_CM-1} 
\end{figure}

\begin{table}
	\caption{Features input sizes and ML model layers for detection inside 
	the operating ($d>1$~m) and the collaborative ($d<1$ m) spaces.}
	\label{tableML} \centering \includegraphics[width=0.48\textwidth]{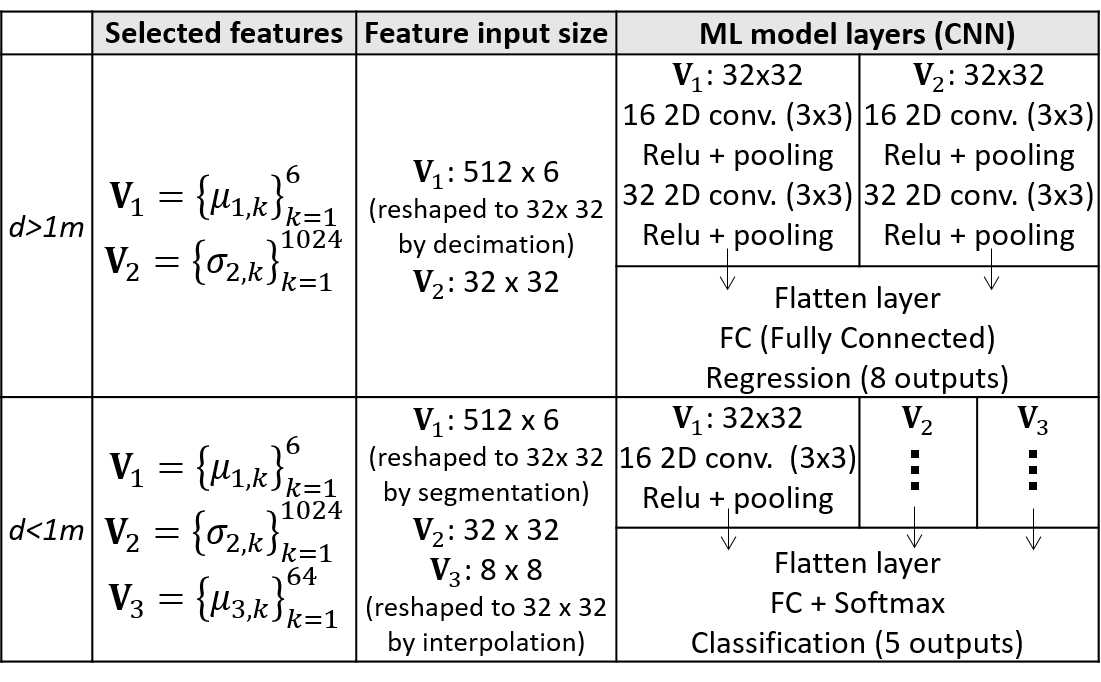} 
\end{table}

\subsection{Worker-robot co-presence monitoring ($d<1$~$\mathrm{m}$)}
\label{subsec:copresence}
In the third use case, we verified the performance of the proposed MDF platform with respect to its ability to recognize the worker in the critical landmarks A, B, C, D and E inside the collaborative space (see Fig.~\ref{fig-layout}.c and~\ref{fig-layout}.d). 
As seen before, the performance analysis is based on the comparison of the confusion matrices to assess precision, recall and average accuracy metrics.
The motion detection system inside the collaborative space must be now more accurate compared with the previous case as monitoring critical co-presence situations. 
Features~$\mathrm{\mathbf{V}}_{1},\mathrm{\mathbf{V}}_{2}$, and~$\mathrm{\mathbf{V}}_{3}$ obtained from the FMCW radars (pipeline $i=1$), the sub-THz detectors ($i=2$) and the IR arrays ($i=3$) are thus fused together.

In Fig.~\ref{sensorfusion_CM-1}, we verified the detection accuracy in the selected positions using the IR array alone (Fig.~\ref{sensorfusion_CM-1}.a)
and by fusing all the features~$\mathrm{\mathbf{V}}_{1},\mathrm{\mathbf{V}}_{2}$,
and~$\mathrm{\mathbf{V}}_{3}$ (Fig.~\ref{sensorfusion_CM-1}.b).
In both cases, we adopt a CNN model with parameters illustrated in Tab.~\ref{tableML} 
(for~$d<1$ m). Similarly as done in the previous section, when fusing all the features
$\mathrm{\mathbf{V}}_{1},\mathrm{\mathbf{V}}_{2}$, and~$\mathrm{\mathbf{V}}_{3}$, the 
$3$ inputs are converted to 2D data structures of similar size. Each 2D structure 
serves as input to the corresponding convolutional filter. Filter outputs are 
flattened being the inputs of a FC layer to classify up to~$6$ HR co-presence 
situations. Data fusion substantially improves the detection performance: considering
the average accuracy, it increases from~$86.9$$\%$ to~$97.5$$\%$.
Similar improvements are observed in terms of precision and recall
for all landmarks.

\section{MDF platform robustness for human--robot collaboration scenarios}

\label{sec:Protective-human--robot-distance}
In this section, we focus more specifically on the analysis of the MDF system robustness. Considering the HRC environment and the related risk mitigation policies, robustness is quantified here in terms of the minimum \emph{protective separation distance} $d_{p}$. 
Distance~$d_{p}$ provides a standardized indicator of the minimum human-robot separation below which the robot must stop or replan its activities. 
Increasing HR separation makes the system more robust to inaccuracies; however, it reduced HRC workspace efficiency. In what follows, we focus on a typical SSM operation scheme (Sect.~\ref{subsec:hrc_cooperation}) where the robot(s) and the operator(s) may move concurrently in the shared workspace.
As shown in Sect.~\ref{subsec:protective}, the protective distance~$d_{p}$ is underpinned by the worker localization uncertainties as well as ML-based detection/classification latencies. In Sect.~\ref{subsec:latency}, localization accuracy and latency of the MDF platform are measured to analyze the 
minimum protective distance for different manipulator speeds. Maximum robot speed and localization uncertainties can be traded to minimize~$d_{p}$ and thus improving system robustness. 

\subsection{Protective human-robot separation distance}

\label{subsec:protective}
The protective human-robot separation distance is defined~\cite{ISOTS15066} as the shortest permissible distance~$d_{p}$ between any moving parts of the robot and any human in the collaborative workspace. The robot is \emph{safety-rated} programmed so that it never gets closer to the operator than~$d_{p}$ by lowering its speed, while if this happens due to operator unsafe motions, the robot system must protectively stop. 
The protective distance depends also on the robot speed~$v_{r}$ and the relative position of the worker.  
It is thus affected by the uncertainty factors related to the detection/classification process. 
Considering the standard~\cite{ISOTS15066}, the general equation for~$d_{p}$, at time~$t_{0}$, can be expressed as

\begin{equation}
\begin{array}{cl}
d_{p}\left(t_{0}\right) & =\int_{t_{0}}^{t_{0}+T_{w}+T_{r}+T_{s}}v_{w}\left(t\right)dt+\int_{t_{0}}^{t_{0}+T_{w}+T_{r}}v_{r}\left(t\right)dt+\\
 & +\int_{t_{0}+T_{w}+T_{r}}^{t_{0}+T_{w}+T_{r}+T_{s}}v_{s}\left(t\right)dt+Z_{r}+Z_{w}
\end{array}\label{eq:ISO-TS-15066}
\end{equation}
where we have indicated the instantaneous velocity terms~$v_{w}\left(t\right)$,
$v_{r}\left(t\right)$ and~$v_{s}\left(t\right)$ (expressed in m/s)
that refer to the directed speed of the worker (moving in the direction
of the robot), the directed speed of the robot (in the direction of
the worker) and directed speed of the robot while stopping, respectively.
$T_{w}$ refers to the worker detection latency (in seconds),~$T_{r}$
indicates the time interval needed to activate the robot stop command while~$T_{s}$ describes the time interval required to completely stop the robot. The worker and robot detection uncertainty factors $Z_{w},T_{w}$ and~$Z_{r}$ correspond to: \emph{i}) the worker localization accuracy~$Z_{w}$ (in meters); \emph{ii}) the worker detection latency~$T_{w}$ (in seconds); and \emph{iii}) the robot localization accuracy $Z_{r}$ (in meters).
In particular, the overall reaction time of the robot system, is the sum of the required time~$T_{w}$ for the detection of the worker, that includes data collection and edge/cloud processing, the activation of robot stop~$T_{r}$, and, finally, the time~$T_{s}$ the robot takes to stop its movements.

In what follows, we assume~$Z_{r}\ll Z_{w}$; in addition, we use the
maximum worker/robot speed values~$v_{w}\leq v_{w}(t)$,~$v_{r}\leq v_{r}(t)$
and the average speed~$v_{s}=\mathrm{E}_{t}[v_{s}(t)]$ of the robot during the stopping phase. By implicitly dropping the current time $t_{0}$, the previous equation can be rewritten as 
\begin{equation}
d_{p}=v_{w}\times(T_{w}+T_{r}+T_{s})+v_{r}\left(T_{w}+T_{r}\right)+v_{s}T_{s}+Z_{w}.\label{eq:dp}
\end{equation}
Therefore,~$d_{p}$ is a function of the worker detection uncertainty factors 
$Z_{w},T_{w}$, while the other parameters are defined in Tab.~\ref{parameters_cell}.
The localization accuracy~$Z_{w}$ and the detection latency~$T_{w}$ are thus
ruled by the proposed MDF platform and are evaluated in the following section.

\subsection{Latency and accuracy analysis from cases studies}

\label{subsec:latency}

In Tab.~\ref{latency}, the accuracy~$Z_{w}$ and the detection latency~$T_{w}$ are measured separately for the detection of worker at distance~$d>1$~m from the robot and the human--robot co-presence at~$d<1$~m. 
The localization accuracy~$Z_{w}$ is obtained for varying landmarks, according to the deployment of Fig.~\ref{fig-layout},
namely the positions~$1,2,...,5$ for~$d>1$~m and positions~$\mathrm{A,B,...,E}$
for~$d<1$ m. The average accuracy is~$Z_{w}=0.54$
m for~$d>1$~m and~$Z_{w}=0.28$ m for~$d<1$~m, respectively.

The latency~$T_{w}$ of the MDF platform corresponds to the time interval between 
the pre-processing of raw data on the edge and the ML based classification of the fused 
features on the cloud. Latency thus depends on the number of pipelines that are needed
by the cloud to process the new location update. More specifically,
the worker localization for~$d>1$~m uses~$2$ pipelines: one consecutive
FFT measurement~$\mathbf{\mathrm{X}}_{k,1}(t)$ obtained from the
$6$ FMCW radars and one~$\mathbf{\mathrm{X}}_{k,2}(t)$ measurement
for each THz detector. The measured latency is~$T_{w}=37$ ms. 
On the contrary, co-presence monitoring  for~$d<1$ m requires larger localization accuracy: the edge now combines~$3$ pipelines, namely~$3$ consecutive FMCW radar measurements~$\mathbf{\mathrm{X}}_{k,1}(t)$, one~$\mathbf{\mathrm{X}}_{k,2}(t)$ measurement for each THz detector and one thermal image~$\mathbf{\mathrm{X}}_{k,3}(t)$.
The need for a larger accuracy causes an increase of the latency to~$T_{w}=90$ ms.

\begin{table}
\caption{Robotic cell parameters for protective human-robot separation distance
$d_{p}$ computation.}
\label{parameters_cell} \centering \includegraphics[width=0.47\textwidth]{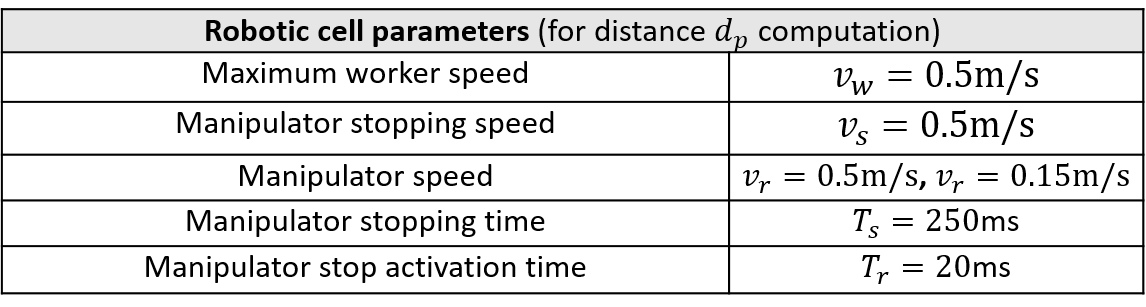} 
\end{table}

\begin{table}
\caption{Analysis of uncertainty factors and protective human-robot separation
distance~$d_{p}$: localization accuracy and latency for worker motion
detection ($d>1$~m as in Fig.~\ref{fig-layout}.b) and worker-robot
co-presence monitoring ($d<1$ m as in Fig.~\ref{fig-layout}.d).}
\label{latency} \centering \includegraphics[width=0.45\textwidth]{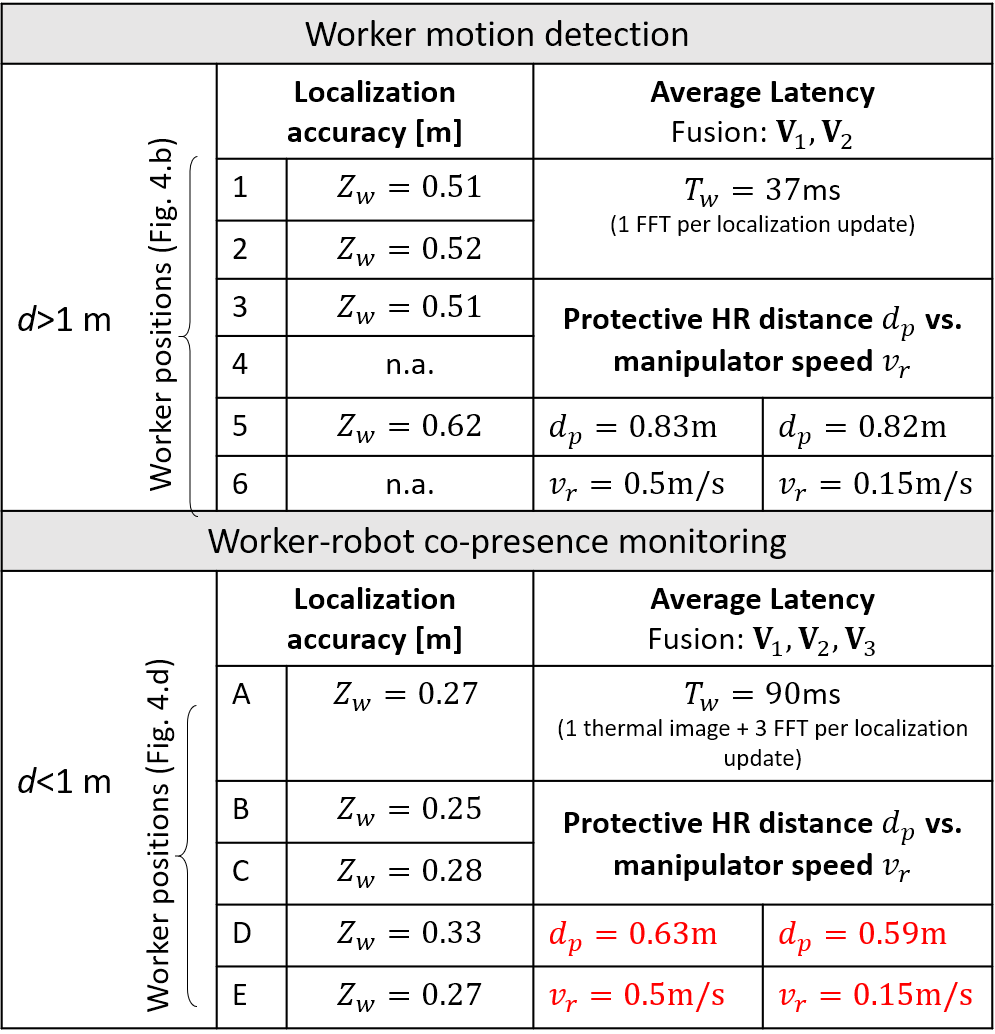} 
\end{table}

Considering the robotic cell parameters in Tab.~\ref{parameters_cell},
and the uncertainties~$Z_{w},T_{w}$, the protective human-robot separation
distance has been computed in the Tab.~\ref{latency} using
(\ref{eq:dp}) for~$d>1$~m and~$d<1$ m, respectively.
Assuming a constant manipulator speed set to a maximum value of~$v_{r}=0.5$
m/s, the separation distance is evaluated as~$d_{p}=0.83$ m for the operator
located in the operating space ($d>1$~m) and as~$d_{p}=0.63$ m
for co-presence monitoring ($d<1$m), respectively. Decreasing the manipulator speed
to~$v_{r}=0.15$ m/s counterbalances the larger reaction time ($T_{w}=90$
ms) for co-presence monitoring ($d<1$ m) as the protective distance now reduces from~$d_{p}=0.63$ m to~$d_{p}=0.59$ m. 
Decreasing robot speed when~$d>1$~m has instead a negligible impact on the protective distance as this is ruled by localization accuracy.

\section{Concluding remarks and open problems}

\label{sec:Conclusions}
We proposed a multisensor data fusion (MDF) edge-cloud platform using radio
signals for opportunistic and anonymous perception of workers inside
a cobot dynamic environment. The MDF platform integrates radio sensing technologies, ranging from infrared, WiFi to sub-THz IoT devices, and is deployed to monitor a fence less human-robot shared workplace, envisioned in Industry 4.0. We proposed a practical solution for the joint analysis of features based on machine learning techniques. In addition, we quantified and discussed the impact of detection uncertainties, namely accuracy and latency, on safe cooperation. The robustness of the MDF system is ruled by the protective human-robot separation distance that we quantified for different human-robot collaborative tasks. Three case studies have been analyzed, including worker counting, motion detection and worker-robot co-presence monitoring. The protective human-robot separation distance takes into account the worker detection latency, accuracy and robotic motions. Based on these results, the proposed computational approach for data manipulation and feature fusion from heterogeneous IoT sensors is promising for the management of advanced human-robot co-presence situations required in next generation industry processes.

% \bibliographystyle{IEEEtran}
% \bibliography{aalto}

\begin{thebibliography}{1}
\bibitem{hrc}P. A. Lasota, T. Fong, and J. A. Shah, ``A survey of
methods for safe human--robot interaction,'' Found. Trends Robot.,
vol. 5, no. 4, pp. 261--349, 2017

\bibitem{hrc-wearable}G. Michalos, N. Kousi, P. Karagiannis, et al.,
``Seamless human robot collaborative assembly -- An automotive case
study,'' Mechatronics, vol. 55, pp. 194-211, ISSN 0957-4158, https://doi.org/10.1016/j,2018.

\bibitem{vic}F. Vicentini, et al., ``Safety Assessment of Collaborative
Robotics Through Automated Formal Verification,'' IEEE Transactions
on Robotics. doi: 10.1109/TRO.2019.2937471, 2019.

\bibitem{hrc2}F. Vicentini, et al., ``Trajectory-dependent safe
distances in human--robot interaction,'' in Proc. IEEE Emerg. Technol.
Factory Autom., pp. 1--4,Sep. 2014.

\bibitem{transf}M. Youssef and F. Kawsar, ``Transformative Computing
and Communication,'' Computer, vol. 52, no. 7, pp. 12-14, July 2019.

\bibitem{hrc-vision}A. Tellaeche, I. Maurtua and A. Ibarguren, ``Use
of machine vision in collaborative robotics: An industrial case.,''
IEEE 21st International Conference on Emerging Technologies and Factory
Automation (ETFA), Berlin, 2016, pp. 1-6. doi: 10.1109/ETFA.2016.7733689

\bibitem{commag_2017}J. Lianghai,et al., "Applying Device-to-Device
Communication to Enhance IoT Services," in IEEE Communications Standards
Magazine, vol. 1, no. 2, pp. 85-91, 2017.

\bibitem{5G-chapter}V. Rampa, et al., ``Opportunistic sensing in
beyond-5G networks: the opportunities of transformative computing,''
in \textit{The 5G Italy Book 2019: a Multiperspective View of 5G},
pp. 461-475, Edited by CNIT, December, 2019, ISBN 9788832170030.

\bibitem{magazine2017} Wu et al. ``Device-Free WiFi Human Sensing:
From Pattern-Based to Model-Based Approaches,'' IEEE Comm. Mag.,
vol. 55, no. 10, 2017.

\bibitem{mag2016} Savazzi et al., ``Device-Free Radio Vision for
assisted living: Leveraging wireless channel quality information for
human sensing,'' IEEE Signal Processing Magazine, vol. 33, no. 2,
2016.

%\bibitem{yous_2017} Yousefi et al., ``A survey on behavior recognition
%using wifi channel state information,'' \emph{IEEE Comm. Mag.}, vol.~55,
%no.~10, 2017.

\bibitem{iot22} Cianca et al., ``Radios as Sensors,'' IEEE IoT
Journal, vol. 4, no. 2, 2017.

\bibitem{sub-TH_2019} S. Kianoush, et al., ``Passive detection and
discrimination of body movements in the sub-THz band: a case study'',
Proc of IEEE Conference on Acoustics, Speech and Signal processing
(ICASSP), 2019.

\bibitem{computer2019}S. Savazzi et al., ``On the Use of Stray Wireless
Signals for Sensing: A Look Beyond 5G for the Next Generation of Industry,''
Computer, vol. 52, no. 7, pp. 25-36, July 2019.

%\bibitem{Shi2016} Shi et al. ``Accurate Location Tracking from CSI-based
%Passive Device-free Probabilistic Fingerprinting,'' IEEE transaction
%on vehicular technology, Doi: 10.1109/TVT.2018.2810307, 2018.

%\bibitem{youseff2013} Seifeldin et al.,``Nuzzer: A large-scale device-free
%localization system for wireless environments,'' IEEE TMC, vol. 12,
%no. 7, 2013.
\bibitem{wang2016human} Wang et al., ``Human respiration detection
with commodity wifi devices: do user location and body orientation
matter'' \emph{ACM Ubicomp}, 2016.

\bibitem{win}Bartoletti et al., ``Device-Free Counting via Wideband
Signals,'' IEEE J. sel. areas Comm., vol. 35, no. 5, 2017.

\bibitem{cellsavazzi2017} S. Savazzi, et al., ``Cellular Data Analytics
for Detection and Discrimination of Body Movements,'' IEEE Access,
vol. 6, pp. 51484-51499, 2018.

\bibitem{IoT_2018} S. Kianoush, et al., "A Cloud-IoT Platform for
Passive Radio Sensing: Challenges and Application Case Studies,"
IEEE Internet of Things Journal, vol. 5, no. 5, pp. 3624-3636, Oct.
2018.

\bibitem{ExManuf_2020} Kong L., et. al.,"Multi-sensor measurement 
and data fusion technology for manufacturing process monitoring: a 
literature review," International Journal of Extreme Manufacturing,
Vol. 2, 022001, 2020.

\bibitem{Sensors_2019} Majumder B.D., et. al.,"Recent advances
in multifunctional sensing technology on a perspective of
multi-sensor system: a review," IEEE Sensors Journal, vol. 19, no. 4,
pp. 1204-1214, Feb. 2019.

\bibitem{IoT2016}S. Kianoush et al., ``Device-Free RF Human Body
Fall Detection and Localization in Industrial Workplaces,'' IEEE
Internet of Things Journal, vol. 4, no. 2, pp. 351-362, April 2017.

%%%%%%% MANUEL-BIBITEM

\bibitem{euroc_book} F. Caccavale, et al., ``Bringing Innovative
Robotic Technologies from Research Labs to Industrial End-users,''
Springer, 2020.

%\bibitem{LEALI201980} F. Leali, F. Pini, and V. Villani, ``Guest
%editorial note: Special issue on human--robot collaboration in industrial
%applications,'', Mechatronics, vol. 58, pp. 80-81, 2019.
\bibitem{HALME2018111} R.-J. Halme, et al.,``{Review of vision-based
safety systems for human--robot collaboration},'' \emph{Procedia
CIRP}, vol.~72, pp. 111--116, 2018.

%\bibitem{MAGRINI2020101846} E.~Magrini, F.~Ferraguti, A.~J. Ronga,
%F.~Pini, A.~D. Luca, and F.~Leali, ``{Human-robot coexistence
%and interaction in open industrial cells},'' \emph{Robotics and
%Computer-Integrated Manufacturing}, vol.~61, p. 101846, 2020. {[}Online{]}.
%Available:
\bibitem{LINSINGER201981} M.~Linsinger et al.,``{Situational task
change of lightweight robots in hybrid assembly systems},'' \emph{Procedia
CIRP}, vol.~81, pp. 81--86, 2019.

\bibitem{LISMONDE20176016} A.~Lismonde, et al.,``Trajectory planning
of soft link robots with improved intrinsic,'' \emph{IFAC-PapersOnLine},
vol.~50, no.~1, pp. 6016--6021, 2017.

\bibitem{ISOTS15066} ``{ISO/TS 15066:2016 Robots and robotic devices
-- Collaborative robots},'' International Organization for Standardization,
Geneva, CH, Standard, 2016.

\bibitem{Vicentini2020} F.~Vicentini, ``{Terminology in safety
of collaborative robotics},'' \emph{Robotics and Computer-Integrated
Manufacturing}, vol.~63, no. December 2019, p. 101921, 2020.

\bibitem{Zanchettin2016} A.~M. Zanchettin, et al., ``{Safety in
Human-Robot Collaborative Manufacturing Environments: Metrics and
Control},'' \emph{IEEE Transactions on Automation Science and Engineering},
2016.

\bibitem{Faroni2019} M.~Faroni, M.~Beschi, and N.~Pedrocchi, ``{An
MPC Framework for Online Motion Planning in Human-Robot Collaborative
Tasks},'' \emph{IEEE International Conference on Emerging Technologies
and Factory Automation, ETFA}, vol. 2019-Septe, pp. 1555--1558, 2019.

\bibitem{Dimostratore} R. Fornasiero, et al.,``{Sustainable Networks
for WEEE Treatment: A Case Study},'' \emph{Procedia CIRP}, vol.~41,
pp. 276--281, 2016.

\bibitem{handbook_reg} S. Chatterjee and J.S. Simonoff, ``{Handbook
of Regression Analysis},'' \emph{ISBN:9780470887165, 2012},

\bibitem{hrc3} N. Pedrocchi, et al.,`` Safe Human-Robot Cooperation
in an Industrial Environment,''International journal on advanced
robotic systems, vol. 10,no.1,2013.

\bibitem{BYNER2019239} C.~Byner, et al.,``{Dynamic speed and separation
monitoring for collaborative robot applications: Concepts and performance},''
\emph{Robotics and Computer-Integrated Manufacturing}, vol.~58, pp.
239--252, 2019.

%\bibitem{wang2017wifall} Wang et al., ``Wifall: Device-free fall
%detection by wireless networks,'' \emph{IEEE TMC}, vol.~16, no.~2,
%2017.

%\bibitem{devicefreeSigg} Sigg et al., ``RF-sensing of activities
%from non-cooperative subjects in device-free recognition systems using
%ambient and local signals,'' IEEE TMC, vol. 13, no. 4, 2014.
%\bibitem{wilson2011} Wilson et al., ``See-through walls: Motion
%tracking using variance-based radio tomography networks,'' IEEE TMC,
%vol. 10, no. 5, 2011.

%\bibitem{Deak2014} Deak et al. `` Detection of multi-occupancy using
%device-free passive localisation,'' IET Wireless Sensor Systems,
%vol. 4, no.3, 2014.

\bibitem{tera_2018}Shur, M."Subterahertz and terahertz sensing of
biological objects and chemical agents", Proc.SPIE, 2018.

\bibitem{AKYILDIZ201416}Akyildiz, I. F. et al.,`` Terahertz band:
Next frontier for wireless communications,'' \emph{Physical Communication.}
, 2014, 12, 16 - 32

\bibitem{tera_magazine} T. Nagatsuma, "Generating millimeter and
terahertz waves," in IEEE Microwave Magazine, vol. 10, no. 4, pp.
64-74, June 2009.

\bibitem{122GHz_radar} S. Scherr, et al., "Miniaturized 122 GHz
ISM band FMCW radar with micrometer accuracy," Proc. of the 2015
European Radar Conference (EuRAD'15), pp. 1-4, Paris, France, Sep.
9-11, 2015.

\bibitem{Short-range FMCW}P. Molchanov, et al., \textquotedbl Short-range
FMCW monopulse radar for hand-gesture sensing,\textquotedbl{} Proc.
of IEEE Radar Conference (RadarCon'15), pp. 1491-1496, Arlington,
USA, May 10-15, 2015.

\bibitem{impatt2007}Mukherjee, M. et al.,"GaN IMPATT diode: a photo-sensitive
high power terahertz source Semiconductor Science and Technology",
2007,vol. 22,no.12,pp.1258, 2007

\bibitem{tera_2015}V.M. Muravev, et al., "Novel Relativistic Plasma
Excitations in a Gated Two-Dimensional Electron System", Phys. Rev.
Lett. 114, 106805, Published 10 March 2015.

%\bibitem{thermal1}Rikke Gade and Thomas B. Moeslund, ``Thermal cameras
%and applications: a survey,'' Machine Vision and Applications, vol.
%25, no. 1, pp. 245--262, Jan 2014.
\bibitem{thermal2}Ash Tyndall, Rachel Cardell-Oliver, and Adrian
Keating, ``Occupancy estimation using a low-pixel count thermal imager,''
IEEE Sensors Journal, vol. 16, no. 10, pp. 3784--3791, May 2016.

\bibitem{thermal3} S. Savazzi, et al.,``Occupancy Pattern Recognition
with Infrared Array Sensors: A Bayesian Approach to Multi-body Tracking,''
IEEE International Conference on Acoustics, Speech and Signal Processing
(ICASSP), Brighton, United Kingdom, pp. 4479-4483, 2019.

\bibitem{palipana2019extracting} S.~Palipana and S.~Sigg, ``Extracting
human context through receiver-end beamforming,'' \emph{IEEE Access},
vol.~7, pp. 154\,535--154\,545, 2019.

\bibitem{sameera2019dfhs} S.~Palipana and S.~Sigg, ``Receiver-side
beamforming to isolate channel perturbations from a human target in
a device-free setting,'' in \emph{The 1st ACM Int. Works. on Device-Free
Human Sensing}, 2019.

\bibitem{palipana2020PerCom} S.~Palipana and S.~Sigg, ``Beamsteering
for training-free recognition of multiple humans performing distinct
activities,'' Proc. of IEEE PerCom, 2020.

\bibitem{thermal4}J. Tanaka, et al., ``Low power wireless human
detector utilizing thermopile infrared array sensor,'' Proc. of IEEE
Sensors, pp. 461-465, Valencia, Spain, Nov 2014.

\bibitem{wot3} Yuyi Mao, et al., ``A Survey on Mobile Edge Computing:
The Communication Perspective,'' IEEE Comm. Surveys and Tutorials,
vol. 19, no. 4, 2017.

\bibitem{wot-sw} L. Mainetti, et. al, ``A Software Architecture
Enabling the Web of Things,'' IEEE Internet of Things Journal, vol.
2, no. 6, pp. 445-454, 2015.

\bibitem{cardoso1993blind} J.-F. Cardoso and A.~Souloumiac, ``Blind
beamforming for non-gaussian signals,'' in \emph{Proc. Radar and
signal process.}, vol. 140, no.~6.\hskip 1em plus 0.5em minus 0.4em\relax
IET, 1993, pp. 362--370.

\bibitem{keogh2001derivative} E.~J. Keogh and M.~J. Pazzani, ``Derivative
dynamic time warping,'' in \emph{Proc. of the Int. Conf. on Data
Mining}.\hskip 1em plus 0.5em minus 0.4em\relax SIAM, 2001.

\bibitem{sharma2012comparison} N.~Sharma \emph{et~al.}, ``Comparison
the various clustering algorithms of weka tools,'' \emph{Facilities},
vol.~4, no.~7, pp. 78--80, 2012.

%\bibitem{Pervasive_Scholz_2011} Scholz et al., ``Sensewaves: Radiowaves
%for context recognition,'' in \emph{Video Proceedings of Pervasive
%2011}, Jun. 2011.
%\bibitem{phy} Holl et al., ``Holography of Wi-Fi Radiation,'' Phys.
%Rev. Lett. 118, 183901, May 2017
%\bibitem{ARTI2017} Kaltiokallio et al., `` ARTI: An Adaptive Radio
%Tomographic Imaging System, '' IEEE TVT, vol. 66, no. 8, 2017.
%\bibitem{woznowski2017sphere} Woznowski et al., ``Sphere: A sensor
%platform for healthcare in a residential environment,'' \emph{Designing,
%Developing, and Facilitating Smart Cities}. Springer, Cham, 2017.
%\bibitem{wang2016gait} Wang et al., ``Gait recognition using wifi
%signals,'' \emph{ACM Ubicomp}, 2016.
%\bibitem{di2016trained} Di~Domenico et al., ``Trained-once device-free
%crowd counting and occupancy estimation using wifi: A doppler spectrum
%based approach,'' \emph{IEEE WiMob}, 2016.
%%\bibitem{liu2014wi}
%%Liu et al., ``Wi-sleep: Contactless sleep monitoring via wifi signals,'' \emph{IEEE RTSS}, 2014.
%\bibitem{liu2015tracking} Liu et al., ``Tracking vital signs during
%sleep leveraging off-the-shelf wifi,'' \emph{ACM MobiHoc}, 2015.
%\bibitem{li2016wifinger} Li et al., ``Wifinger: talk to your smart
%devices with finger-grained gesture,'' \emph{ACM Ubicomp}, 2016.
%\bibitem{wang2016we} Wang et al., ``We can hear you with wi-fi!''
%\emph{IEEE TMC}, vol.~15, no.~11, 2016.
%\bibitem{feng2017mais} Feng et al, ``Mais: Multiple activity identification
%system using channel state information of wifi signals,'' in \emph{WASA},
%2017.
%\bibitem{zhao2016emotion} Zhao et al., ``Emotion recognition using
%wireless signals,'' \emph{ACM Mobicom}, 2016.
%%\bibitem{DeviceFreeRecognition_Sigg_2013}
%%Sigg et al., ``Rf-based device-free recognition of simultaneously conducted activities,'' \emph{ACM Ubicomp Adjunct}, 2013.
%%\bibitem{Muneeba_2017_Geospatial}
%%aja et al., ``Towards pervasive geospatial affect perception,'' \emph{Springer GeoInformatica}, 2017. 
%\bibitem{cloud_complexity2017}C.Yang et al., ``Sensing-Data Curation
%for the Cloud is Coming: A Promise of Scalable Cloud-Data-Center Mitigation
%for Next-Generation IoT and Wireless Sensor Networks,'' IEEE Consumer
%Electronics Magazine, Vo.6, no.4, pp.48-56, 2017.
%\bibitem{IoT2017}Yasin et al.,`` Ultra-Low Power, Secure IoT Platform
%for Predicting Cardiovascular Diseases,'' IEEE TCS, vol.64, no. 9,
%2017.
%%\bibitem{Raja_2016_CoSDEO}
%%Raja et al., ``Applicability of rf-based methods for emotion recognition: A survey,'' in \emph{IEEE PerCom adjunct}, 2016.
%%\bibitem{Pervasive_Sigg_2014b}
%%Sigg et al., ``Teach your wifi-device: Recognise gestures and simultaneous activities from time-domain rf-features,'' \emph{IJACI}, vol.~6, no.~1, 2014.
%%\bibitem{shi2012activity}
%%Shi et al., ``Activity recognition from radio frequency data: Multi-stage recognition and features,'' in \emph{IEEE VTC Fall}, 2012.
%\bibitem{Edgecaching2017} Liang et al.,``Enhancing QoE-Aware Wireless
%Edge Caching With Wireless SDN,'' \emph{IEEE TWC}, vol. 16, no. 10,
%2017.
%\bibitem{Zhu2017} Zhu, et. al, ``Robust and passive motion detection
%with COTS WiFi devices,'' IEEE Tsinghua Science and Technology J.,vol.22,
%no.4, 2017.
%\bibitem{Factcloud2014} Savazzi et al., ``Wireless Cloud Networks
%for the Factory of Things: Connectivity modeling and layout design,''
%IEEE IoT J., vol.1, no.2, 2014.
%\bibitem{Sense_cloud} Alamri, et al.,``A Survey on Sensor-Cloud:
%Architecture, Applications, and Approaches, '' Int. J. of Distr.
%Sensor Netw., vol.9, no. 2, 2013.
%\bibitem{IFCIoT2017} Munir et al., ``Integrated Fog Cloud IoT: A
%novel architectural paradigm for future IoT, '' IEEE Cons. Electr.
%Mag., vol.6, no. 3, 2017.
%\bibitem{LQI2017}Jayasri et al.,``Link Quality Estimation for Adaptive
%Data Streaming in WSN,'' Int. J. Wirel. Personal Comm., vol.94, no.3,
%2017.
%\bibitem{Chang2017}Chang et al.,``FitLoc: Fine-grained and low-cost
%device-free localization for multiple targets over various areas,''
%IEEE/ACM TN, 2017.
%%\bibitem{ICC2016}
%%Kianoush et al.,``Pre-deployment performance assessment of device-free radio localization systems,'' IEEE ICC, 2016.
%\bibitem{Patel2013} Pu et al., ``Whole-home gesture recognition
%using wireless signals, '' Mobicom, 2013.
%\bibitem{yousef2014} Saeed et al.,``Ichnaea: A low-overhead robust
%WLAN device-free passive localization system,'' IEEE J. Sel. Top.
%SP, vol. 8, no. 1, 2014.
%%\bibitem{shi2012activity} Shi et al., ``Activity recognition from
%%radio frequency data: Multi-stage recognition and features,'' in
%%\emph{IEEE VTC Fall}, 2012.
%\bibitem{CSIloc2016} Fang et al., ``Channel State Reconstruction
%Using Multilevel Discrete Wavelet Transform for Improved Fingerprinting-Based
%Indoor Localization,'' IEEE Sensors J., Vol.16, no. 21, 2016.
%%\bibitem{ICC2017}
%%Kianoush et al., ``Tracking of frequency selectivity for device-free detection of multiple targets,'' IEEE ICC Workshops, 2017.
%\bibitem{semantic} Alaya et al.,``Toward semantic interoperability
%in oneM2M architecture,'' IEEE Comm. Mag., vol. 53, no. 12, 2015.
%%\bibitem{wang2016interacting}
%%Wang et al., ``Interacting with soli: Exploring fine-grained dynamic gesture recognition in the
%%  radio-frequency spectrum,'' \emph{ACM UIST}, 2016.
%\bibitem{singh2017smart} Singh et al., ``Smart city environmental
%perception from ambient cellular signals,'' in \emph{International
%Conference on Algorithms and Architectures for Parallel Processing}.
%Springer, 2017.
%%\bibitem{yao2015rf}
%%Yao et al., ``Rf-care: Device-free posture recognition for elderly people using a passive rfid tag array,'' \emph{Mobiquitous}, 2015.
%\bibitem{zou2017grfid} Zou et al., ``Grfid: A device-free rfid-based
%gesture recognition system,'' \emph{IEEE TMC}, vol.~16, no.~2,
%2017.
%\bibitem{lien2016soli} Lien et al., ``Soli: Ubiquitous gesture sensing
%with millimeter wave radar,'' \emph{ACM TOG}, vol.~35, no.~4, 2016.
%\bibitem{adib2015capturing} Adib et al., ``Capturing the human figure
%through a wall,'' \emph{ACM TOG}, vol.~34, no.~6, 2015.
%\bibitem{wot} D. Guinard, et al. ``Towards the web of things: web
%mashups for embedded devices,'' Proc. of the International World
%Wide Web Conference 2009 (WWW 2009), Madrid, Spain, April 2009.
%\bibitem{wot2} He Li, et al., ``Learning IoT in Edge: Deep Learning
%for the Internet of Things with Edge Computing,'' IEEE Network, vol.
%32, no. 1, pp. 96-101, 2018.
%\bibitem{paas}A. Ferrer, et. al. ``Multi-cloud Platform-as-a-service
%Model, Functionalities and Approaches,`` Procedia Computer Science.
%97. pp. 63-72, October, 2016.
%\bibitem{personalCloud} Wu et al., ``Composable 10: A Novel Resource
%Sharing Platform in Personal Clouds'', CloudCom, 2009.
%\bibitem{PCAtipping} M. Tipping, C. Bishop, ``Probabilistic principal
%component analysis,'' J. of Royal Stat. Society Series, col 21 no.
%3, pp. 611-622, 1999.
%\bibitem{PCA2002} R. Vidal, "Subspace Clustering," IEEE Signal
%Processing Magazine, vol. 28, no. 2, pp. 52-68, March 2011.
%\bibitem{m2m_semantic} Swetina et al., ``Toward a Standardized Common
%m2m Service Layer Platform: Introduction to onem2m,'' IEEE Wirel.
%Comm., vol. 21, no 3, 2014.
%\bibitem{json}K. Maeda, ``Performance evaluation of object serialization
%libraries in XML, JSON and binary formats,'' Proc. of Digital Information
%and Communication Technology and it's Applications (DICTAP), pp. 177-182,
%2012.
%\bibitem{Gerla} Gerla, ``Vehicular Cloud Computing'', Med-Hoc-Net,
%2012.
%\bibitem{VANET} Whaiduzzaman et al., ``A survey on vehicular cloud
%computing,'' J. Netw. \& Comp.Applications, vol 40, 2014.
%\bibitem{Burgoon} Burgoon et al., ``Nonverbal communication''.
%Allyn \& Bacon, Boston, MA.2016.
%\bibitem{change} Basseville et al., ``Detection of Abrupt Changes'',
%Theory and Applications. Prentice-Hall, 1993.
%%\bibitem{SVM}
%%Vapnik, ``Statistical learning theory'', Wiley, New York, 1998.

%\bibitem{csitool}Linux 802.11n CSI tool, http://dhalperi.github.io/linux-80211ncsitool/\#
%publications,2012.
\end{thebibliography}

\end{document}